\documentclass[review,12pt,authoryear]{elsarticle}

\journal{Computers and Geotechnics}

\usepackage{amsmath}
\usepackage{amssymb}
\usepackage{graphicx}
\usepackage{booktabs}
\usepackage{array}
\usepackage{enumitem}
\usepackage{lineno}
\usepackage{url}
\usepackage{hyperref}
\usepackage{soul}
\usepackage{tabularx}
\usepackage{makecell}
\graphicspath{{./}}
\newcolumntype{Y}{>{\raggedright\arraybackslash}X}
\newcolumntype{C}{>{\centering\arraybackslash}p}

\biboptions{authoryear,round}

\begin{document}

\begin{frontmatter}

\title{Permeability and microcrack geometry: Dynamic loading induced evolution}

\author[llnl]{Rigoberto Moncada\corref{cor1}}
\ead{moncadalopez1@llnl.gov}
\cortext[cor1]{Corresponding author}
\author[llnl]{Efrem Vitali}

\affiliation[llnl]{organization={Lawrence Livermore National Laboratory},
            addressline={7000 East Avenue},
            city={Livermore},
            postcode={94550},
            state={CA},
            country={USA}}

\begin{abstract}
We propose a novel method for modeling permeability evolution in rocks under dynamic loading conditions. Accurate representation of permeability is essential for extraction and containment applications of fluids within geological strata. The main complexity of this representation is capturing the permeability response under variable loading conditions. Low-porosity brittle rocks are highly sensitive to pore geometry and connectivity changes, which can cause drastic permeability and anisotropy alterations. Moreover, limited dynamic permeability measurements and material heterogeneity increase modeling challenges.

This work offers a crack-geometry-based permeability evolution model for high strain-rate conditions that accounts for microcracking. Rock voids are represented by an evolving network of penny-shaped cracks with prescribed orientations. Changes in permeability are described by adjustments of crack geometric microvariables, including aperture, length, and distance. The model considers crack opening and closure, as well as fracture-energy-driven crack propagation. Its main novelty is the introduction of formulations for plastic-strain-driven nucleation, coalescence after full connectivity is reached, and strain-rate-dependent fracture toughening. The network's geometric response to dynamic loading affects void connectivity, fluid conductance, and preferential flow paths.

We analyze how different loading conditions affect permeability through the evolution of crack geometry. The formulation is implemented in the GeoDyn hydrocode and evaluated under tension, shear, compression, and symmetric impact loading. The study shows a strong correlation between permeability and crack length, with propagation being a key factor. It also demonstrates how permeability increase can be inhibited by poor connectivity and crack closure. The model captures permeability changes of several orders of magnitude depending on stress state, strain rate, crack orientation, and connectivity, consistent with observations in brittle rocks under high strain-rate conditions.
\end{abstract}

\begin{keyword}
permeability evolution \sep microcracks \sep dynamic loading \sep brittle rocks \sep computational geomechanics 
\end{keyword}

\end{frontmatter}


\section{Introduction}\label{intro}

In this work, we focus on the evolution of absolute permeability. Absolute, or intrinsic, permeability of a material links the pressure gradient and flow rate to a material's pore or crack fabric. This property applies to saturated conditions and a single fluid phase. Changes in pore microstructure caused by mechanical, thermal, and chemical processes \citep{Peng2021} can temporarily or permanently modify void geometry and how well these voids are connected, altering intrinsic permeability. 

Our main interest is the effect mechanical loading has on permeability variations, including elastic and plastic deformations, and fracture. Quasistatic loading is considered, but our focus is on the dynamic regime. In this paper we will cover low-porosity brittle rocks as the foundation for this method. Thus, the most relevant pores contributing to permeability are microcracks and fractures resulting from nucleation and propagation. In these materials, flow is affected more by crack permeability than matrix permeability.

Isotropic compression tends to reduce void cross-sectional area and diminish porosity and permeability \citep{Morris2003}. Most evolution models, such as Kozeny--Carman \citep{Ma2015}, use the stress state or effective confining pressure to predict permeability changes. They relate permeability to porosity via power-law or exponential relations. However, porosity change is not the only mechanism through which pore structure can be modified. Permeability of a material is linked to connectivity or percolation \citep{Peach1996}, or how well pores communicate with each other. Proportionality between porosity and permeability may not hold depending on pore connectivity, and permeability may vary while porosity remains relatively constant \citep{Payton2022}. Hence, materials subject to fracture also undergo connectivity rise from increased crack density and length \citep{Perol2016}, not only porosity growth.  For high strain rates, geometry changes such as pore crushing and dilation due to plastic failure and damage \citep{Zhu1997, Morrow2001} are also vital. While damage has multiple denotations, our definition is ``volume fraction of cracks in a solid material.'' Damage grows by the nucleation of new cracks or the enlargement of existing cracks. Moreover, directionality of crack networks \citep{Oda1985} also affects permeability response significantly. As a result, permeability formulations can be improved by expressing damage as a tensor \citep{Liao2023, Zhang2023}, linking loading conditions to directional material properties.

Understanding the evolution of permeability is critical for applications involving flow in porous materials subjected to changing loading conditions. So far, quasistatic conditions have been widely researched. Quasistatic studies include permeability, or flow, decrease for containment applications such as carbon capture \citep{Zhang2024b} and underground storage of methane, hydrocarbons, and radioactive waste \citep{UliaszMisiak2024}. Other steady state work is centered in increasing permeability via preferential paths, like hydraulic fracturing, geothermal energy generation, leach mining, and hydrocarbon production \citep{Gehne2019}. However, there is less work on permeability changes driven by dynamic loading. Mining and construction projects with geomaterials subject to high strain rates due to blasting have to account for failure mechanisms that can facilitate flow via fracturing or reduce it via pore crushing \citep{Zhao2021}. Similarly, nuclear explosion monitoring techniques that rely on permeability predictions to estimate radionuclide detection \citep{Bourret2019, Heath2021} must consider enhancement or inhibition of underground flow. Given the growing interest and limited work for these conditions, we prioritize dynamic loading in this paper.

Modeling permeability evolution of brittle materials subject to high strain-rate conditions is a multi-faceted problem \citep{Fuchs2025, Li2025}. Models should account for strain rate dependency, connectivity changes, and large variations of permeability. We introduce an analytical model capable of representing these behaviors using crack geometry microvariables. In this work, damage is defined by crack fabric, which can be modified through mechanical loading and subsequently mapped into permeability. Crack microstructure evolution is driven by stress, fracture propagation, and coalescence. This work is based on scalar permeability, dynamic loading, and penny-shaped crack models \citep{Gueguen1989, Peach1996, Zhu1999, Simpson2001, Zuo2006, Paliwal2008, Perol2016}, and can be combined with the directionality proposed by \citep{Oda1985}. Our model accounts for reversible aperture opening and closure, irreversible crack propagation, and aperture-to-length aspect ratio conservation. Its main novel contribution is a crack-geometry-based permeability framework that introduces formulations for plastic-strain-driven nucleation, coalescence after the full-connectivity threshold is reached, and strain-rate-dependent fracture toughening. In this study, we present the scalar form of the model to isolate the role of crack geometry under dynamic loading; tensorial permeability and evolving crack orientations are left for future work.

This paper is structured as follows. First, in the ``Background'' section, we review empirical, semi-empirical, and theoretical relations for permeability evolution. We then define permeability in terms of crack geometric variables. Next, the ``Model'' section describes the main features of the formulation. These features include stress-driven crack aperture evolution, dynamic crack length propagation, and crack distance reduction through nucleation. The ``Simulations and Results'' section introduces a series of simplified loading conditions. It includes a ``single-element'' section that evaluates the effects of individual model components. The ``multiple-elements'' section then presents a more general impact uniaxial test examining permeability evolution in space and time. Both simulation sections define porosity as the sum of crack and matrix components, with the latter represented by a pseudocap strength model \citep{Vorobiev2021} that includes porosity compaction as implemented in the GeoDyn hydrocode. We finalize the paper with a ``Discussion'' section of our results, followed by the ``Conclusions'' section. This work will be extended in a companion paper focusing on the tensorial aspects of our formulation.

\section{Background} \label{backg}

\subsection{Permeability definition}

At its most basic level, permeability is the material property that permits fluid flow under a pressure gradient or energy potential. Permeability represents intrinsic pore geometry controlling flow and it is independent from fluid density, temperature, or viscosity. Hence, the definition of intrinsic, or absolute, permeability for single-phase flow is obtained from Darcy's law:

\begin{equation}
v = -\frac{1}{\mu} k (\Delta p + \rho g)
\end{equation}

\noindent where $v$ is the effective flow velocity along the fluid pressure gradient $\Delta p$, $\rho$ is the fluid density, $g$ is the acceleration due to gravity, $\mu$ is the dynamic fluid viscosity, and $k$ is the intrinsic permeability. Permeability has units of area, reflecting its geometric relationship to the surface area of pores and cracks.

A more general expression \citep{bear1972}, using index notation and allowing permeability to be represented as a tensor, is :

\begin{equation} \label{fullK}
\overline{v_i} = -\frac{\overline{B T_{ij}}}{\mu} \left( 
\frac{\partial \bar{p}}{\partial x_j} + \rho g \frac{\partial z}{\partial x_j}
\right) =
-\frac{k_{ij}}{\phi \mu} \frac{\partial p}{\partial x_j}
\end{equation}

\noindent where the overbar indicates a spatial average of the measured quantity, $x_j$ denotes the spatial coordinate, $z$ is the height of the fluid column used to measure hydrostatic pressure, $\bar{p}$ is the hydraulic pressure, and $p$ is the combination of the two ($p = \bar{p} + \rho g z $). The remaining terms define the permeability tensor as $k_{ij} = \phi \overline{B T_{ij}}$, where each component represents an aspect of pore or crack geometry:

\begin{itemize}
  \item $\phi$ -- Porosity: fraction of the total volume occupied by voids.
  \item $B$ -- Conductance: measure of flow rate through the pores of a material, controlled by geometry like pore diameter, throat radius, or crack aperture. It represents the microstructural equivalent of the Hagen--Poiseuille relation \citep{Zhang_2014}.
  \item $T_{ij}$ -- Non-random porous-medium operator: a measure of connectivity or tortuosity of crack networks with respect to directions $i$ and $j$, where $i$ denotes the flow direction and $j$ denotes the direction of the pressure gradient.
\end{itemize}

The above formulation explicitly shows the roles of porosity, connectivity, orientation, and flow conductance. As such, it provides a useful reference for developing analytical models in which permeability evolution arises as a stress-driven response of crack geometry. 

\begin{figure}[ht!] \centering \noindent\includegraphics[width=10cm]{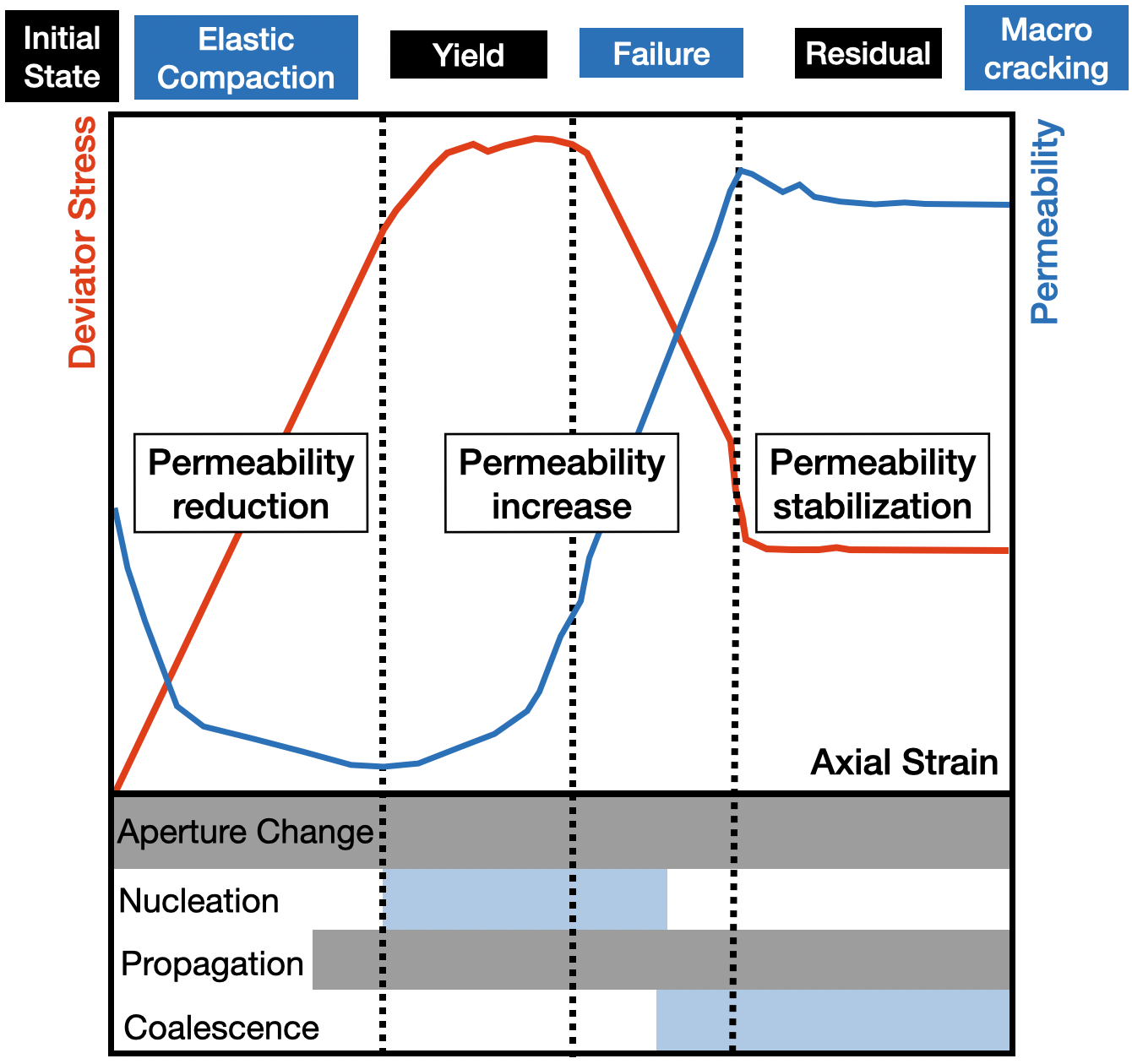} \noindent\includegraphics[width=9cm]{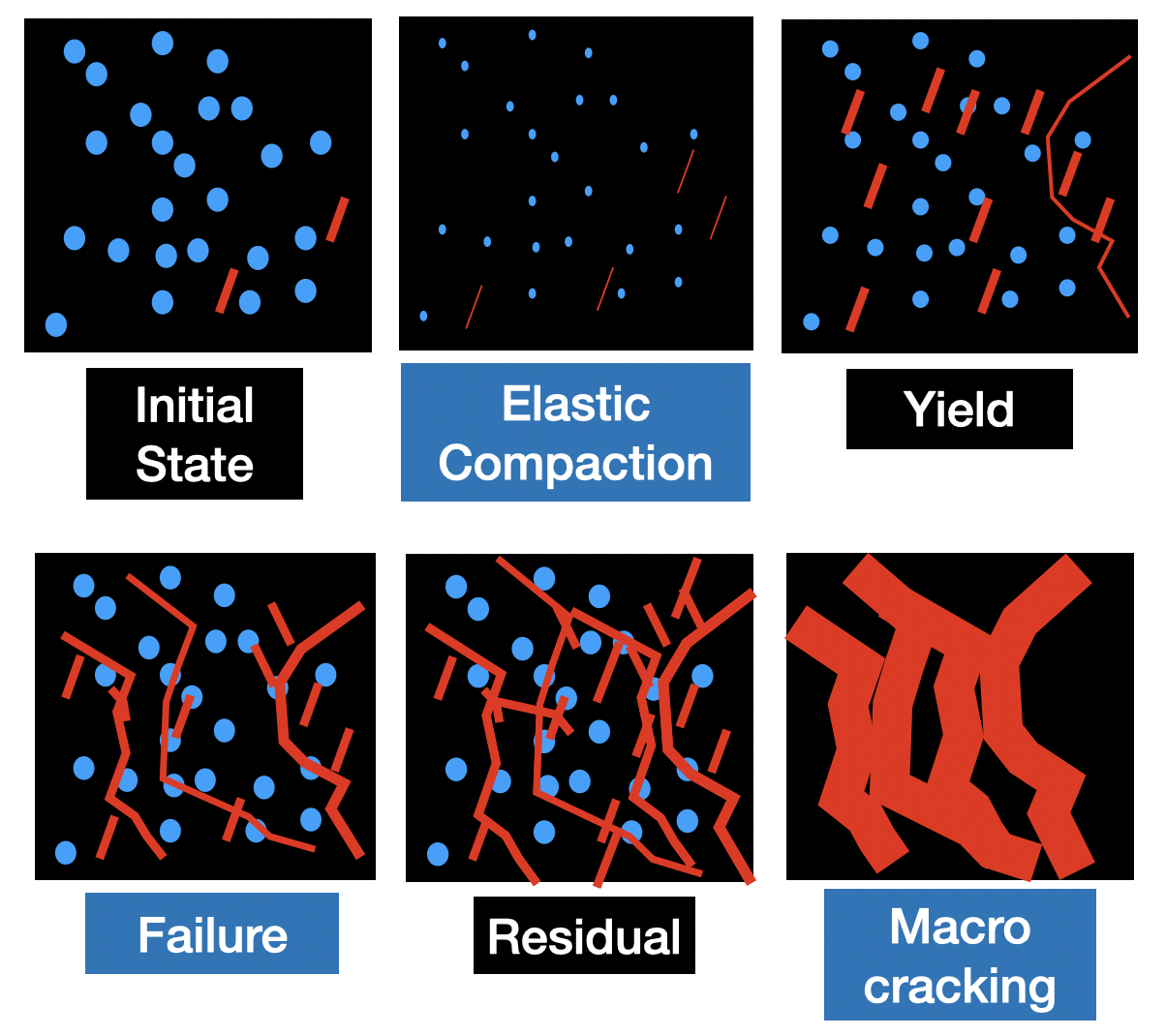} \caption{Permeability evolution under axial strain for a triaxial test, defined in terms of pore and crack microstructure changes. (Based on results by \citep{Zhang2022})} \label{evo_stages} \end{figure}

The evolution of permeability is commonly measured by a triaxial compression tests \citep{Mitchell2008, Wang2021, Zhang2022, Sueyoshi2023}. In such tests, a background confining pressure is applied while strain is controlled along a principal axis, and permeability is measured along that direction. As deformation progresses, loading is often paused at prescribed strain intervals to measure permeability. Figure~\ref{evo_stages} shows the deviator stress and permeability as a function of axial strain during a triaxial test \citep{Zhang2022}. It illustrates that permeability changes are directly linked to loading-induced modifications of void geometry. These changes are qualitatively divided into distinct regimes. For initially unloaded materials, the first stage is `Elastic Compaction'. During compaction, compressive stresses close voids and permeability decreases. This reduction continues until a minimum permeability is reached during the elastic regime, when further elastic deformation cannot reduce pore volume. Within this elastic regime, minor crack nucleation and propagation may occur, leading to a limited increase in permeability. After `Yield', crack nucleation and propagation intensify, and pore dilation increases permeability. As deformation progresses toward failure, crack growth leads to full crack network connectivity increasing permeability. Upon `Failure', crack coalescence induces larger crack apertures that result in additional rise in permeability. Once the material reaches a `Residual' post-failure state, crack growth ceases and permeability stabilizes. Another implication is that the system of microcracks evolved into a `Macroscopic Fracture'. Transition to this scale is beyond the scope of our study.

This permeability sequence corresponds to quasistatic triaxial loading for brittle materials. Highly dynamic loading or ductile material responses lead to different trends, including permeability increases of several orders of magnitude beyond quasistatic changes \citep{Mitchell2022, Meyer2024}. Moreover, dynamic conditions may create residual pore crushing states where permeability approaches zero \citep{Morris2003}. These studies suggest that permeability evolution models would benefit from a clear relationship between stress evolution and changes in pore-crack geometry.

\subsection{Approaches to permeability evolution} \label{cont_disc}

Different formulations have been developed to predict permeability evolutions, some more compatible with changes in connectivity, orientation or porosity. One of the most common empirical relations for permeability is the Kozeny--Carman model, in which permeability is expressed as a function of porosity $\phi$, mean grain (or crystal) diameter $\bar{d}$, and a shape and grain-size distribution factor $C_{\mathrm{sd}}$, through a power-law relation \citep{Ma2015, DazCuriel2022}:

\begin{equation}
      k = C_{\mathrm{sd}} \frac{\phi^3}{(1-\phi)^2} \bar{d}^2.
\end{equation}

In terms of permeability evolution or response to loading conditions, continuum porosity-dependent models are the most common \citep{Ma2015}. Most of these models are formulated for the quasistatic regime and have been validated under such experimental conditions. Reviews describing permeability evolution models for porous materials, such as coal, have been compiled by \citep{Pan2012, Zhang2014, Gao2022} and demonstrate how continuum approaches can be modified to estimate permeability.

In addition to porosity-based formulations, there are effective stress- or pressure-based models \citep{Ma2015}. These models are generally applicable to isotropic loading regimes and elastic deformations. Other approaches incorporate dilation of pore structure due to plastic yielding \citep{Zhang2016}. In such formulations, a scalar or tensorial damage variable is evolved using a yield-surface (e.g., Tresca, von Mises, or Drucker--Prager) and it is related to permeability changes \citep{Zhang2022}. Damage tensor is interpreted as a measure of crack volume density and orientation within a material \citep{Lubarda1993}. Some damage-based models are able to represent permeability evolution beyond yield and represent changes due to fracturing and material failure \citep{Zhang2016}.

An alternative approach to permeability evolution is based on discrete representations of void space, such as pore network models (PNMs) \citep{Zhao2020, Luo2023} and discrete fracture networks (DFNs) \citep{Thomas2020, Pietruszczak2024}. These methods allow permeability to evolve via geometric changes of voids without continuum parametrizations and account for connectivity changes. Nonetheless, they are limited in properly representing the wide variability of void shape, size, and orientation due to computational constraints and scarce experimental or in-situ data of pore structures \citep{Hosseinzadegan2023}. To address these challenges, some authors \citep{dienes1978statistical, Dienes2006} have proposed Statistical Crack Mechanics as a means of homogenizing the effects of discrete crack populations. In this formulation, pore or crack microstructure is described statistically using a set of microvariables, which are then used to upscale permeability behavior at the continuum scale. Changes in these microvariables are directly related to changes in permeability, allowing prediction of isotropic permeability for random crack distributions \citep{Gueguen1989}. Microvariable-based permeability models \citep{Peach1996, Simpson2001, Simpson2003, Benson2006} assume idealized pore or crack geometries and have been used to predict scalar permeability evolution in response to loading conditions. 

Based on our findings, we propose using statistical crack mechanics. This approach combines the advantages of continuum and discrete formulations to describe pore and crack networks through a reduced set of geometric microvariables. Building on this concept, we develop an analytical permeability evolution model that explicitly links dynamic stress-driven crack geometry changes to permeability.

\subsection{Permeability evolution for high strain rates}

Our primary objective is to model permeability evolution in brittle, low-porosity rocks under high strain-rate loading. Consequently, pore microstructure-based approaches that include dynamic fracture mechanics are fundamental. They have been used to identify changes in pore geometry under high strain rates and to predict significant increases in permeability magnitude \citep{Perol2016}. High strain-rate loading is typically associated with sharp increases in permeability prior to material failure. Dynamic fracture processes promote faster connectivity and crack coalescence, which lead to increased flow.

Several works \citep{Zuo2006, Paliwal2008, Perol2016} employ idealized winged penny-shaped crack geometries to define dynamic stress intensity factors or fracture energy criteria that govern crack growth. Loading conditions induce crack geometry evolution that is responsible for permeability changes, reproducing the drastic permeability increases observed in experimental measurements \citep{Aben2020}. A major restriction of these models is that they are exclusively scalar and do not account for crack orientation; therefore, they cannot capture crack anisotropy. However, orientation plays a critical role in crack opening and propagation. It is also worth noting that due to experimental challenges and the difference between the timescales of dynamic loading and flow measurements, permeability evolution data under dynamic conditions is often limited to initial and final states. Intermediate measurements are scarce, and tracking the full temporal evolution of permeability remains challenging \citep{inproceedingsLi2024}.  While analytical models are effective in bridging data gaps, the cited work lacks anisotropy, nucleation, coalescence, and strain rate dependent crack growth capabilities. These limitations are addressed in this work.

The model presented in the next section follows the works cited above by employing crack aperture and length changes. In addition, it incorporates nucleation, coalescence changes, and strain rate dependent parameters for dynamic loading conditions. The model also includes the orientation of penny-shaped cracks and their surface stresses into the evolution equation. As a result, our formulation improves permeability predictions and identifies preferential flow paths.

\section{Model} \label{model}

\subsection{Geometric Definition of Permeability}

Building on the statistical crack mechanics discussed in Section~\ref{cont_disc}, we return to a homogenized, microstructure-based approach for defining permeability. To express permeability in terms of pore microstructure, the heterogeneous void space is approximated using idealized geometries. Pore microstructure can be represented by spheres, elongated cylinders, flat pennies, or combinations thereof. In this work, we focus on low-porosity crystalline rocks, such as granite, diorite, and rock salt. For these materials, penny-shaped cracks are widely considered representative of the voids controlling flow \citep{Peach1996}.

\begin{figure}[ht!]  
\centering 
\noindent \includegraphics[width=\textwidth]{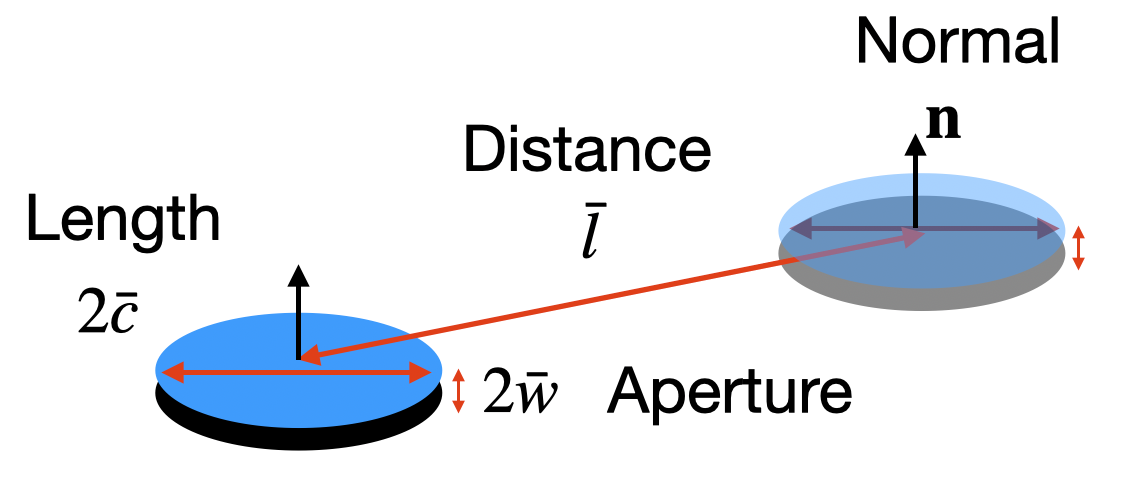}
\caption{Main features of a penny-shaped crack.} \label{penny_geo}
\end{figure}

The scalar permeability of a medium containing randomly oriented penny-shaped cracks \citep{Gueguen1989, Simpson2001} is expressed as
\begin{equation} \label{geo_k}  
k = \frac{4\pi}{15} f(\bar{c}, \bar{l}) \frac{\bar{w}^3 \bar{c}^2}{\bar{l}^3},
\end{equation}

\noindent where $\bar{c}$ is the representative crack length, $\bar{l}$ is the representative distance between cracks, $\bar{w}$ is the representative crack aperture, and $f$ is the connectivity factor, ranging from 0 to 1. Based on Figure~\ref{penny_geo}, length is defined as the radius and aperture as the half-thickness of the penny. Representative values refer to weighted averages. These averages ensure that the effective permeability of the penny-shaped crack ensemble matches that of the real material. This representation requires knowledge of crack size distributions of aperture, length, and distance. This approximation allows moderate crack size variability but is most reliable for near-uniform distributions.

Figure~\ref{penny_geo} introduces an additional key variable: the crack normal vector $\mathbf{n}$, which introduces directional effects. The percolation or connectivity factor $f$ depends on crack length and distance and can be derived using a Bethe lattice approximation, in which the coordination number of penny-shaped cracks is $Z=4$ \citep{Peach1996, Torquato_2002-ax}. Under this assumption, the percolation factor proposed by \cite{Peach1996} represents the fraction of geometric features belonging to the percolating or connected cluster and is given by
\begin{equation}\label{percol_eq}
 f = 1 - 4\left[ \left( \frac{1}{p^\mathrm{int}} - \frac{3}{4} \right)^{1/2} - 0.5 \right]^3 
     + 3\left[ \left( \frac{1}{p^\mathrm{int}} - \frac{3}{4} \right)^{1/2} - 0.5 \right]^4,
\end{equation}

\noindent where $p^\mathrm{int}$ is the intersection probability of penny-shaped cracks based on their geometric properties,
\begin{equation}\label{inter_eq}
 p^\mathrm{int} = \frac{\pi^2 \bar{c}^3}{4\,\bar{l}^3}.
\end{equation}

For convenience, permeability may also be expressed in terms of crack porosity $\phi_c$ as
\begin{equation} \label{include_eq}
k = \frac{2 f(\bar{c}, \bar{l}) \bar{w}^2 \phi_t}{15}, ~\text{where} ~\phi_t = \phi_c = 2 \pi \bar{c}^2 \bar{w} / \bar{l}^3,
\end{equation}

\noindent for a unit volume and $\phi_t$ represents total porosity. Introducing crack porosity explicitly is advantageous when material models track matrix and crack porosity separately at the microscale. In Equation~\ref{include_eq} total porosity $\phi_t$ can be modified to also include matrix porosity $\phi_m$ by defining it as
\begin{equation} \label{tot_poro}
\phi_t = \phi_c~+~\phi_m.
\end{equation}
\noindent But for the low-porosity crystalline rocks considered, crack porosity constitutes the dominant contribution to permeability. This homogenized formulation provides the geometric basis for mapping stress-driven crack evolution into permeability changes.

\begin{figure}[ht!]  
\centering 
\noindent \includegraphics[width=\textwidth]{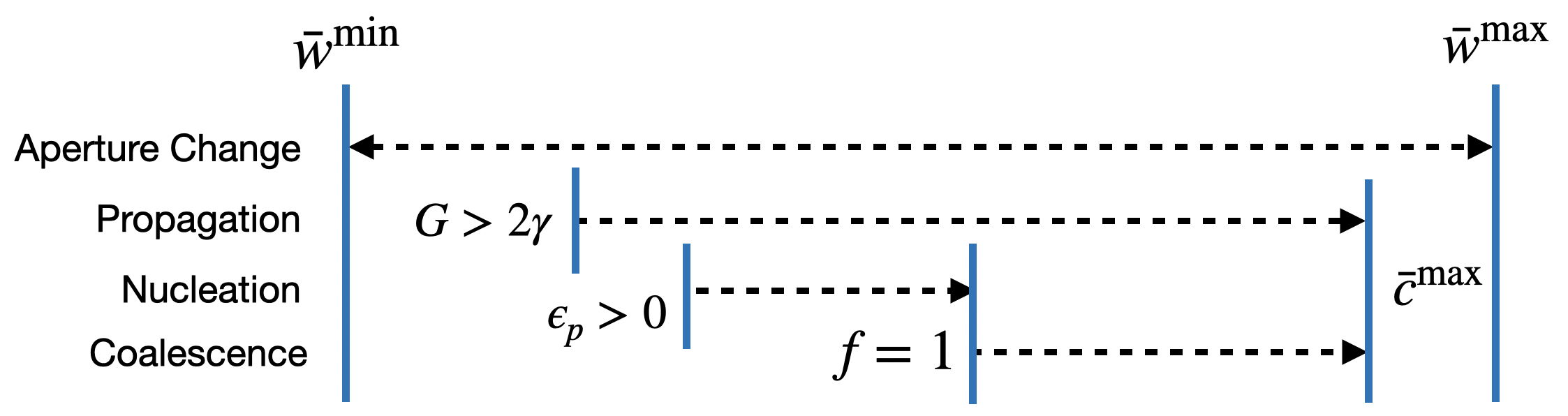}
\caption{Lower and upper bounds within the model for crack geometry changes. Double-headed arrow indicates reversibility. Single-headed arrows indicate irreversibility. Minimum and maximum allowable apertures are represented by $\bar{w}^\mathrm{min}$ and $\bar{w}^\mathrm{max}$. Maximum allowable length is defined as $\bar{c}^\mathrm{max}$. $G$ stands for energy release rate, $\gamma$ is critical surface energy and $\epsilon_p$ is plastic strain.} \label{maxmin}
\end{figure}

The evolution of the permeability equation depends on the evolution of microvariables $\bar{w}$, $\bar{c}$ and $\bar{l}$. The evolution of each geometric microvariable is bounded by specific physical constraints: aperture roughness and element size limits for $\bar{w}$, fracture propagation and element size limits for $\bar{c}$, and nucleation-controlled bounds for $\bar{l}$. These constraints are illustrated in Figure~\ref{maxmin} and discussed in detail in the subsequent sections.

\noindent Consequently, permeability can be predicted analytically if the stress-induced evolution of crack geometry is accurately tracked. With our formulation, we can make permeability evolution more tractable by expressing geometric changes not only in terms of porosity $\phi_c(\bar{w}, \bar{c}, \bar{l})$, but also considering aperture (cubic law) $\bar{w}$ and connectivity (percolation) effects $f(\bar{c}, \bar{l})$ \citep{bear1972}.

Given the complexity of modeling crack nucleation explicitly, we adopt a simplified plastic-strain-based model to predict nucleation and update crack distance $\bar{l}$. More comprehensive mechanical models are applied to the representative crack length $\bar{c}$ and aperture $\bar{w}$, which evolve through fracture propagation and opening or closure, respectively. Together, these three geometric microvariables govern the evolution of crack porosity $\phi_c$, thus, the accuracy of the proposed permeability model relies on the fidelity of the poro-elasto-plastic constitutive relations and fracture mechanics models utilized. 

In the following sections, we introduce the mechanical models governing the evolution of crack aperture, length, and distance for penny-shaped cracks. For simplicity, we consider only one family, or group of cracks with the same geometry, to describe the ensemble of cracks. Additionally, no new crack families are formed over time. We also assume a constant normal vector. 

\subsection{Crack aperture changes}

For a linear elastic response, the current crack aperture $\bar{w}^{\mathrm{el}}$ is computed analytically as
\begin{equation} \label{elastic_opening}
\bar{w}^{\mathrm{el}} = \max \left( \bar{w}^o - \bar{c}\,\beta(\boldsymbol{\sigma}), \; \bar{w}^{\mathrm{min}} \right),
\end{equation}

\noindent where $\bar{w}^{\mathrm{min}}$ is the minimum allowable crack aperture, which depends on intrinsic material roughness, and $\bar{w}^o$ is the initial aperture. The term $\beta$ is a normalized pressure term accounting for the aperture-to-length ratio of the crack. Following \cite{Gavrilenko1989}, $\beta$ is defined as
\begin{equation}
\beta(\boldsymbol{\sigma}) = -\frac{2 (1-\nu^2)\,\sigma^n}{E},
\end{equation}

\noindent where $E$ is Young’s modulus, $\nu$ is Poisson’s ratio, and $\sigma^n$ is the projection of the stress tensor $\boldsymbol{\sigma}$ onto the crack normal (refer to \textit{Supplementary Material} section for this definition).

Eq.~\eqref{elastic_opening} can be generalized to account for both elastic and inelastic deformation by replacing the stress-based expression with the total strain tensor. More generally, the current crack aperture $\bar{w}$ along a given normal vector $\mathbf{n}$ is determined by the projection of the total strain tensor $\boldsymbol{\epsilon}$ as
\begin{equation} \label{wstrain}
\bar{w} = \max \left( \left( 1 + \mathbf{n} \cdot \boldsymbol{\epsilon} \mathbf{n} \right)\bar{w}^o, \; \bar{w}^{\mathrm{min}} \right).
\end{equation}

An important experimental and numerical observation for cracks is that they tend to preserve their aperture-to-length aspect ratio $A$ during growth, typically within a power-law range between 0.5 and 2.0 relative to their initial proportions \citep{Mayrhofer2019}. This assumption yields
\begin{equation} \label{aspect_ratio}
A = \frac{\bar{c}^o}{\bar{w}^o} = \frac{\bar{c}}{\bar{w}},
\end{equation}

\noindent such that Eq.~\eqref{elastic_opening} is reformulated as
\begin{equation} \label{elastic_opening_ar}
\bar{w}^{\mathrm{el}} = \max \left( \frac{\bar{c}}{A} - \bar{c}\,\beta(\boldsymbol{\sigma}), \; \bar{w}^{\mathrm{min}} \right),
\end{equation}

\noindent and Eq.~\eqref{wstrain} becomes
\begin{equation} \label{w_general}
\bar{w} = \max \left( \left( 1 + \mathbf{n} \cdot \boldsymbol{\epsilon} \mathbf{n} \right)\frac{\bar{c}}{A}, \; \bar{w}^{\mathrm{min}} \right).
\end{equation}

When crack propagation becomes extensive, overlapping cracks may coalesce, resulting in additional aperture increase. In this work, coalescence is assumed to occur once the connectivity factor reaches unity ($f=1$). For simplicity, we consider the overlap between exactly two cracks. When $f=1$, the current aperture $\bar{w}$ is replaced by a coalescence-modified aperture $\bar{w}^{\mathrm{coal}}$ given by
\begin{equation}
\bar{w}^{\mathrm{coal}} =
\frac{ \bar{c}^{\mathrm{per}} + g^f \left( \bar{c} - \bar{c}^{\mathrm{per}} \right) }{\bar{c}} \, \bar{w},
\end{equation}

\noindent where $\bar{c}^{\mathrm{per}}$ is the crack length at the percolation threshold ($f=1$) and $g^f = 2$ is a growth factor. This growth factor is chosen under the assumption that the maximum aperture of two identical cracks coalescing is twice their aperture. As an upper bound, we define the maximum possible aperture $\bar{w}^{\mathrm{max}} = \bar{c}^{\mathrm{max}}/A$. Aperture changes are a mix of reversible and irreversible processes because these changes result from elastic strain, inelastic strain, aperture-to-length aspect ratio conservation, and coalescence. All other geometric changes are treated as irreversible. The next section introduces the crack propagation mechanisms and the associated fracture criteria.

\subsection{Crack length propagation}

Crack length increase depends on the energy available for fracture under multiple modes. The representative crack length is updated as
\begin{equation}
\bar{c} = \min \left( \bar{c}^{o} + \Delta \bar{c}, \; \bar{c}^{\mathrm{max}} \right),
\end{equation}

\noindent where $\bar{c}$ is the current crack length and $\bar{c}^{\mathrm{max}}$ is an upper bound related to the characteristic element size $L$. Conservatively, we define $\bar{c}^{\mathrm{max}} = L/4$, to prevent microcrack length from propagating beyond the element. A different criterion for limiting crack growth would be to calculate the maximum admissible crack porosity prior macroscopic fracturing. However, element size is adopted as the default limit.

Current crack length is determined by the initial crack length $\bar{c}^o$ and the crack length increment $\Delta \bar{c}$. The magnitude of this increment depends on loading conditions and material mechanical response. In this work, we focus on elastic microcrack crack propagation, which typically occurs in brittle low porosity crystalline rocks, such as granite. Stress intensity factors and fracture toughness relations for quasistatic elasto-plastic and dynamic elastic loading regimes are detailed in the Supporting Information section.

Crack growth is constrained as irreversible. A reduction in crack driving energy halts further propagation, such that crack growth resumes only if an energy state higher than previously imposed is reached. Crack length reduction is not considered because chemical and thermal healing mechanisms are beyond the scope of this work.

Transitions between quasistatic and dynamic regimes affect the crack driving energy and, therefore, influence the percolation or connectivity. Dynamic fracture regimes increase crack driving energy and may significantly alter permeability compared to quasistatic loading. To identify this transition, we first define the average strain rate within the material as
\begin{equation}
\dot{\bar{\epsilon}} =
\frac{1}{9}
\sqrt{
\dot{\epsilon}_{11}^2 +
\dot{\epsilon}_{22}^2 +
\dot{\epsilon}_{33}^2 +
2 \dot{\epsilon}_{12}^2 +
2 \dot{\epsilon}_{23}^2 +
2 \dot{\epsilon}_{13}^2
}.
\end{equation}

When the average strain rate exceeds a dynamic threshold $\dot{\bar{\epsilon}}_{\mathrm{dyn}}$, typically in the range $0.01$--$0.1~\mathrm{s^{-1}}$, the material is considered to be in the dynamic regime. Under these conditions, crack propagation depends on stress intensity factors and the crack driving energy or release rate. For dynamically propagating cracks, we adopt the Griffith instability criterion as used by \cite{Zuo2006, Perol2016}, such that crack growth occurs only if the energy release rate $G$ exceeds twice the surface fracture energy $\gamma$:
\begin{equation}
G(\boldsymbol{\sigma}, \mathbf{n}, \bar{c}) \ge 2\gamma.
\end{equation}

Depending on the loading regime, a general expression for the crack length increment is
\begin{equation} \label{dyn_deltaC}
\Delta \bar{c} =
\begin{cases}
\Delta \bar{c}_{\mathrm{SS}}, & \dot{\bar{\epsilon}} < \dot{\bar{\epsilon}}_{\mathrm{dyn}}, \\
\Delta \bar{c}_{\mathrm{dyn}}, & \dot{\bar{\epsilon}} \ge \dot{\bar{\epsilon}}_{\mathrm{dyn}},
\end{cases}
\end{equation}

\noindent where $\Delta \bar{c}_{\mathrm{SS}}$ denotes the quasistatic crack increment and $\Delta \bar{c}_{\mathrm{dyn}}$ the dynamic crack increment. A fully dynamic plastic formulation becomes necessary as the material approaches the brittle--ductile transition under increasing confining pressure and temperature. However, for the low-porosity crystalline rocks and the moderate confining pressures considered, crack growth response remains predominantly elastic even under dynamic loading. Extensions to ductile regimes are a subject of future work. Details on the strain rate dependent formulations are found in the \textit{Supplementary Material} section.

\subsection{Crack nucleation}

Mechanisms governing crack nucleation are inherently complex and involve dislocation processes at the crystal scale \citep{Wu2025}.  Plastic strain, particularly during the early stages of material failure, leads to nucleation \citep{Zhang2016b}. Crack nucleation leads to increased crack density and reduced distance between cracks. This densification typically stabilizes when the material reaches full connectivity or percolation, then crack propagation and aperture opening dominate permeability evolution \citep{Roy2021}.

Based on these observations, crack nucleation in this model is active only when the connectivity factor is less than unity ($f < 1$) and plastic strain is present ($\epsilon_p > 0$). We define crack density as the number of microcracks, $\bar{N}$, per unit volume, with nucleation corresponding to an increase in $\bar{N}$. We assume the following inverse relationship between crack density $\bar{N}$, and crack distance $\bar{l}$:

\begin{equation}
\bar{N} = \frac{1}{\bar{l}}.
\end{equation}
\noindent Crack densification is then linked to the accumulated plastic strain through
\begin{equation}
\bar{N} = \bar{N}^o \left( 1 + \epsilon_p \right)^{n_c},
\end{equation}
\noindent which is equivalently expressed as
\begin{equation} \label{ldens}
\bar{l} = \frac{\bar{l}^o}{\left( 1 + \epsilon_p \right)^{n_c}}.
\end{equation}

\noindent The sensitivity of nucleation to plastic strain is regulated by the nucleation exponent $n_c$; for simplicity, we assume $n_c = 1$. After full connectivity is attained, propagation is the preferred failure mode and coalescence is activated. 

Among geometric microvariables, crack length is the most dominant in our model. Crack distance and aperture are tied to crack length. Length increase determines aperture changes due to aperture-to-length aspect ratio conservation and coalescence. Crack length affects distance by controlling nucleation through the connectivity limit $f=1$ \citep{Deng2022}. The next section presents results for the model, highlighting the impact of each geometric microvariable and showing the dominance of crack length. The \textit{Supplementary Material} section has additional details on the relationship between our model and basic 1D permeability expressions; this sections also summarizes the steps that define initial crack geometry from existing crack size distributions imported from microscopy or CT scan data.

\section{Simulations and Results}

We evaluate the permeability model using a series of controlled loading conditions designed to demonstrate how different crack geometric microvariables affect permeability evolution. The tests focus on the roles of crack aperture, crack length, and crack distance under distinct loading regimes. In particular, we examine how strain rate, crack propagation, crack closure, and crack distance reduction modify permeability.

The mechanical response is computed using a material model that includes elastic deformation, plastic strain accumulation, strain-rate dependency, matrix porosity evolution, pseudocap yield failure, and spall-induced strength degradation \citep{Vorobiev2021}. This formulation is appropriate for brittle geomaterials subjected to strain rates between $0.1$ and $1000~\mathrm{s^{-1}}$. GeoDyn \citep{osti_928156} provides the computational platform used to exercise this model. The mechanical properties and initial crack geometry are summarized in Table~\ref{simpleT}; additional granite parameters are taken from \cite{Vorobiev2021}.

\begin{table}[htbp]
    \centering
    \caption{Material properties for granite used in tests}
    \label{simpleT}
    \scriptsize
    \setlength{\tabcolsep}{3pt}
    \renewcommand{\arraystretch}{1.15}
    \begin{tabularx}{\textwidth}{>{\centering\arraybackslash}p{0.20\textwidth} Y >{\centering\arraybackslash}p{0.17\textwidth} >{\centering\arraybackslash}p{0.13\textwidth}}
    \toprule
    \textbf{Variable} & \textbf{Meaning} & \textbf{Value} & \textbf{Units} \\
    \midrule
    ${\phi_c}^\mathrm{init}$ & Initial crack porosity & 0.015 & - \\
    ${\phi}^\mathrm{init}$ & Initial matrix porosity, single-element tests & 0.01 & - \\
    ${\phi}^\mathrm{init}$ & Initial matrix porosity, multiple-elements test & 0.01 & - \\
    $E$ & Young's modulus & 25 & GPa \\
    $\nu$ & Poisson ratio & 0.28 & - \\
    $S$ & Shear modulus & 2.3 & GPa \\
    $\rho$ & Density & 2707 & kg m$^{-3}$ \\
    $\dot{\bar{\epsilon}}_{\mathrm{dyn}}$ & Dynamic strain rate & 0.03 & s$^{-1}$ \\
    $\dot{\bar{\epsilon}}_{\mathrm{tough}}$ & Toughness strain rate threshold & 0.03 & s$^{-1}$ \\
    $\mu$ & Friction coefficient & 0.65 & - \\
    $a_{\mathrm{Transverse}}$ & Transverse wave speed & 3000 & m s$^{-1}$ \\
    $a_{\mathrm{Rayleigh}}$ & Rayleigh wave speed & 2770 & m s$^{-1}$ \\
    $\gamma$ & Critical fracture surface energy & 500 & J m$^{-2}$ \\
    $K^\mathrm{Ic}$ & Mode I toughness & 1.75 & MPa m$^{0.5}$ \\
    $K^\mathrm{IIc}$ & Mode II toughness & 2.33 & MPa m$^{0.5}$ \\
    $K^\mathrm{IIIc}$ & Mode III toughness & 1.05 & MPa m$^{0.5}$ \\
    $Y_c$ & Compressive strength & 0.069 & GPa \\
    $Y_t$ & Tensile strength & 0.01 & GPa \\
    $\bar{c}$ & Crack length & 0.20 & mm \\
    $\bar{w}$ & Crack aperture & 0.005 & mm \\
    $\bar{l}$ & Crack distance & 0.35 & mm \\
    $\bar{w}^\textrm{min}$ & Minimum crack aperture & 1e-6 & mm \\ 
    $\mathbf{n}$ & Normal vector, single-element tests & $(0,0,1)$ & - \\
    $\mathbf{n}$ & Normal vector, multiple-elements test & $(0.707,0.707,0)$ & - \\
    \bottomrule
    \end{tabularx}
\end{table}

\subsection{Single-element 1D setup and results}

Single-element simulations are used to isolate the response of the permeability model under idealized loading paths. Low, intermediate, and high (dynamic) strain rates (Table~\ref{loadT}) are applied to uniaxial tension, simple shear, and triaxial compression tests. Simulation durations are adjusted so that each strain-rate test reaches the same final strain, with lower strain rates requiring a larger number of time steps.

For all single-element simulations, the crack normal vector is aligned with the $z$-axis. Therefore, the crack lies in the $xy$-plane and the crack normal is parallel to the principal loading direction in the tension and compression cases. Permeability is measured in the $x$-direction. We track the evolution of normalized crack aperture, crack length, crack distance, and permeability, where each normalized quantity is divided by its initial value before loading.

\begin{table}[htbp]
    \centering
    \caption{Loading conditions for single-element material library simulations}
    \label{loadT}
    \scriptsize
    \setlength{\tabcolsep}{3pt}
    \renewcommand{\arraystretch}{1.2}
    \begin{tabularx}{\textwidth}{>{\raggedright\arraybackslash}p{0.24\textwidth} >{\centering\arraybackslash}p{0.13\textwidth} >{\centering\arraybackslash}p{0.20\textwidth} Y}
    \toprule
    \textbf{Loading scenario} & \textbf{dt ($\mu$s)} & \textbf{Number of steps} & \textbf{Strain rate (s$^{-1}$)} \\
    \midrule
    Uniaxial tension 
    & 1 
    & 5000, 1000, 500 
    & $\dot{\epsilon}_{zz}=1,\,5,\,10$ \\
    Simple shear 
    & 1 
    & 10000, 4000, 500 
    & $\dot{\gamma}_{xz}=0.4,\,10,\,80$ \\
    Triaxial compression 
    & 1 
    & 5000, 1250, 500 
    & $\dot{\epsilon}_{zz}=-2,\,-8,\,-20$; $\dot{\epsilon}_{xx}=\dot{\epsilon}_{yy}=-1,\,-4,\,-10$ \\
    \bottomrule
    \end{tabularx}
\end{table}

\subsubsection{Uniaxial tension}

\begin{figure}[ht!]  
\centering 
\noindent\includegraphics[width=\textwidth]{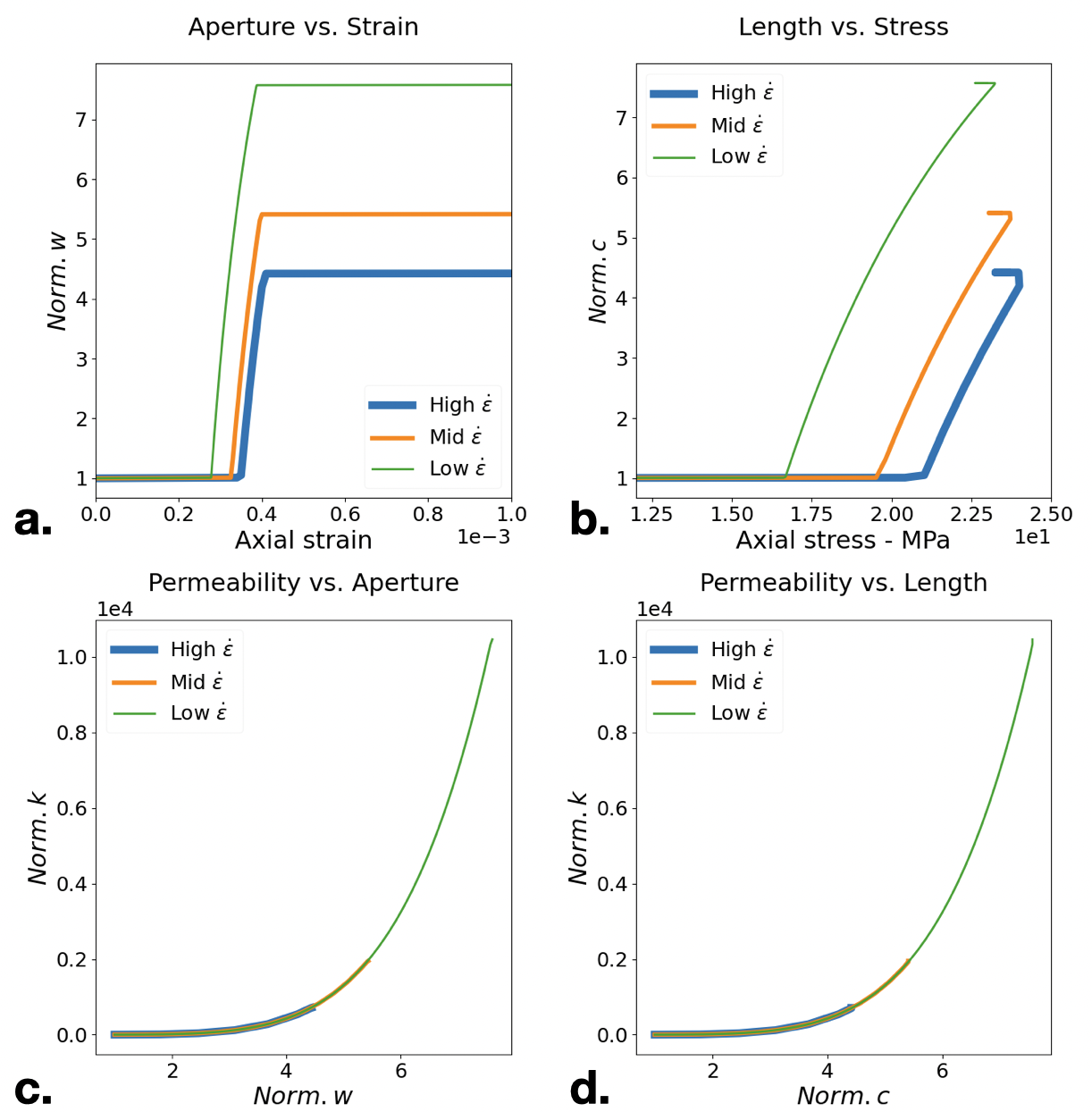}
\caption{Stress to Crack Geometry to Permeability mapping under Uniaxial Tension for different strain rates: a. aperture vs axial strain, b. length vs axial stress, c. permeability vs aperture, and d. permeability vs length. All crack geometries and permeability are normalized with respect to their initial values. Crack distance results do not vary.}  \label{simple_tension}
\end{figure}

Uniaxial tension is applied in the $z$-direction at the three strain rates listed in Table~\ref{loadT} to exercise Mode I fracture. Figure~\ref{simple_tension}a shows the normalized crack aperture as a function of axial strain. In all cases, aperture remains nearly constant until a threshold strain is reached, after which it increases abruptly. This threshold occurs earlier for lower strain rates, and the final aperture is also larger at lower strain rates. The sharp aperture increase is primarily controlled by crack propagation explained in Eq.~\textit{SM}-23. Because all three loading rates are within the dynamic regime, the strain-rate-dependent critical surface energy in Eq.~\textit{SM}-27 controls propagation speed and, therefore, the strain/stress threshold. Conservation of the aperture-to-length aspect ratio (Eqs.~\ref{aspect_ratio}--\ref{w_general}) controls aperture tensile response. 

Figure~\ref{simple_tension}b shows a similar behavior as Figure~\ref{simple_tension}a. Crack length increase under tension is described by Eq.~\textit{SM}-19. As shown in Eq.~\textit{SM}-27, lower strain rates produce less toughening, allowing propagation to begin at a lower stress and resulting in greater crack length. Once failure occurs, stress decreases and propagation stops. Length plateaus in Figure~\ref{simple_tension}b reflect this arrest, while also showing the irreversibility of crack growth: crack length does not decrease after stress release.

Figures~\ref{simple_tension}c and \ref{simple_tension}d show the resulting permeability as a function of aperture and length, respectively. Tensile loading produces permeability increases of about three to four orders of magnitude with respect to the initial value. This rise occurs primarily due to crack propagation, aperture increase, coalescence, and the onset of full connectivity. In these figures it is noticeable that the maximum permeability for the high strain rate case is inferior to that of lower strain rate cases. This difference is caused by strain-rate-induced dynamic toughening, leading to cracks that are shorter and with smaller apertures. 

Crack distance is not included in Figure~\ref{simple_tension} because it does not change appreciably in the tensile case. Due to rapid length increase, full connectivity is reached very quickly and nucleation-driven distance reduction is inhibited in the model (see Figure~\ref{maxmin}).

\subsubsection{Simple shear}

\begin{figure}[ht!]  
\centering 
\noindent\includegraphics[width=\textwidth]{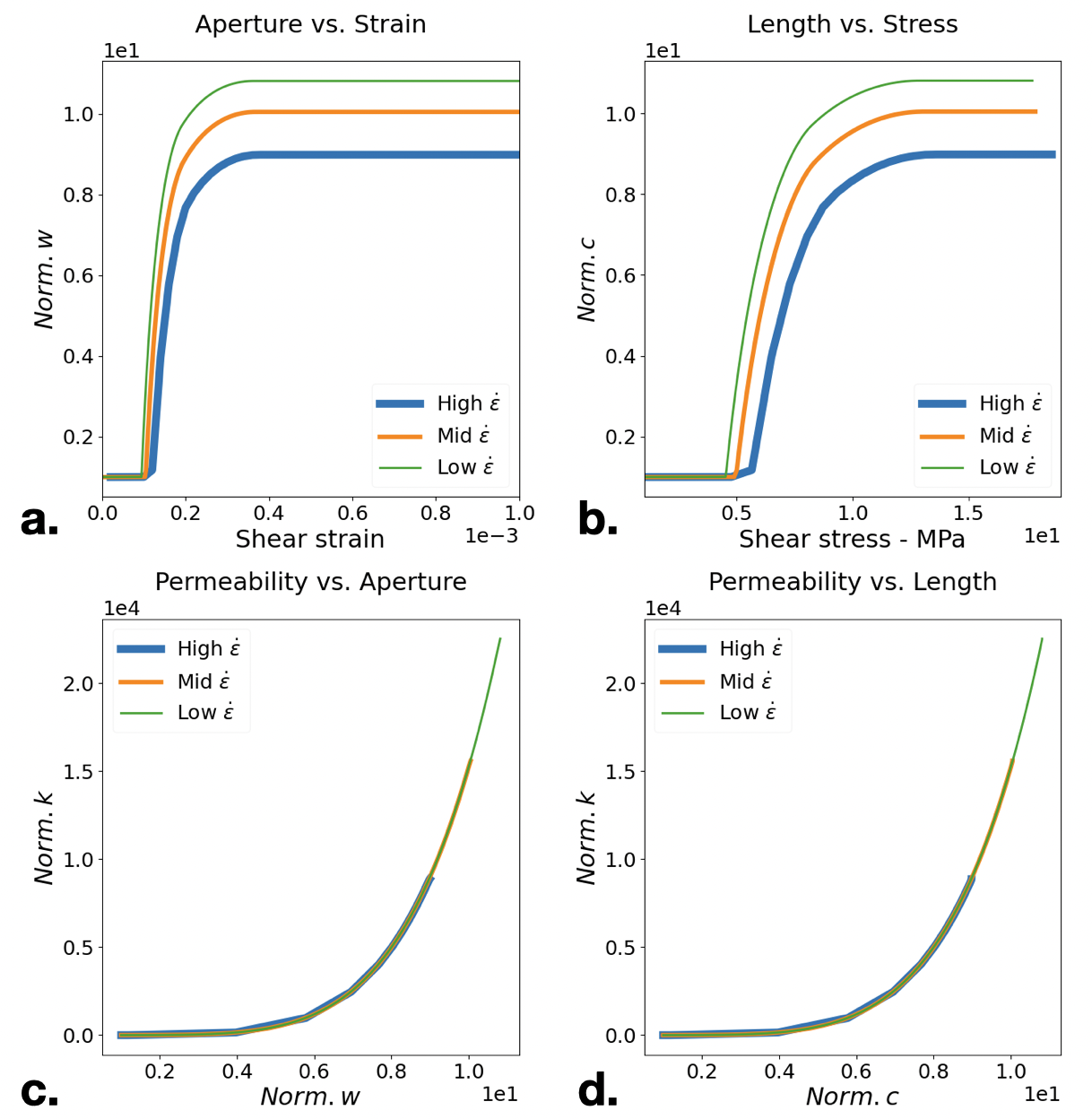}
\caption{Stress to Crack Geometry to Permeability mapping under Simple Shear for different strain rates: a. aperture vs shear strain, b. length vs shear stress, c. permeability vs aperture, and d. permeability vs length. All crack geometries and permeability are normalized with respect to their initial values. Crack distance results do not vary.}  \label{simple_shear}
\end{figure}

Simple shear is applied in the $xz$-direction for the three shear strain rates listed in Table~\ref{loadT} to exercise Mode II fracture. Figure~\ref{simple_shear}a shows aperture evolution as a function of shear strain. For shear, as in the tensile case, aperture increases drastically; however, it then stabilizes more gradually compared to uniaxial tension. Also, in this shear case, aperture increase is controlled by the constant aperture-to-length aspect ratio as described by Eqs.~\ref{aspect_ratio}--\ref{w_general}.

Figure~\ref{simple_shear}b shows crack propagation as a function of shear strain under different strain rates. Propagation is driven by shear fracture, as explained by Eq.~\textit{SM}-21. In these cases crack growth is arrested when reaching maximum length within the element, as explained in Figure~\ref{maxmin}. This figure shows that lower strain rates produce greater crack growth; this growth initiates at lower shear stresses because strain-rate-dependent toughening is weaker, like in the uniaxial tension test (Eq.~\textit{SM}-27). 

Figures~\ref{simple_shear}c and \ref{simple_shear}d show that simple shear increases permeability by approximately three to four orders of magnitude with respect to the initial values. Like in tension, this increase results from the combined effects of crack propagation, aperture rise, coalescence, and full connectivity. Similar to the previous section, for the same target strain, low strain rates reach higher final permeabilities than high strain rates. This is due to the greater toughening generated at higher strain rates, which inhibit crack growth (Eq.~\textit{SM}-27). 

Another similarity with tension is that full connectivity is reached rapidly and crack distance does not change. This is the reason we do not show distance evolution for Figure~\ref{simple_shear}. 

\subsubsection{Triaxial compression}

Triaxial compression is applied at different strain rates (Table~\ref{loadT}) while maintaining a fixed ratio between the lateral compressive strain rates in the $x$- and $y$-directions and the axial compressive strain rate in the $z$-direction. Since compression does not promote crack propagation, we focus on crack closure and nucleation-driven crack distance reduction.

\begin{figure}[ht!]  
\centering 
\noindent\includegraphics[width=\textwidth]{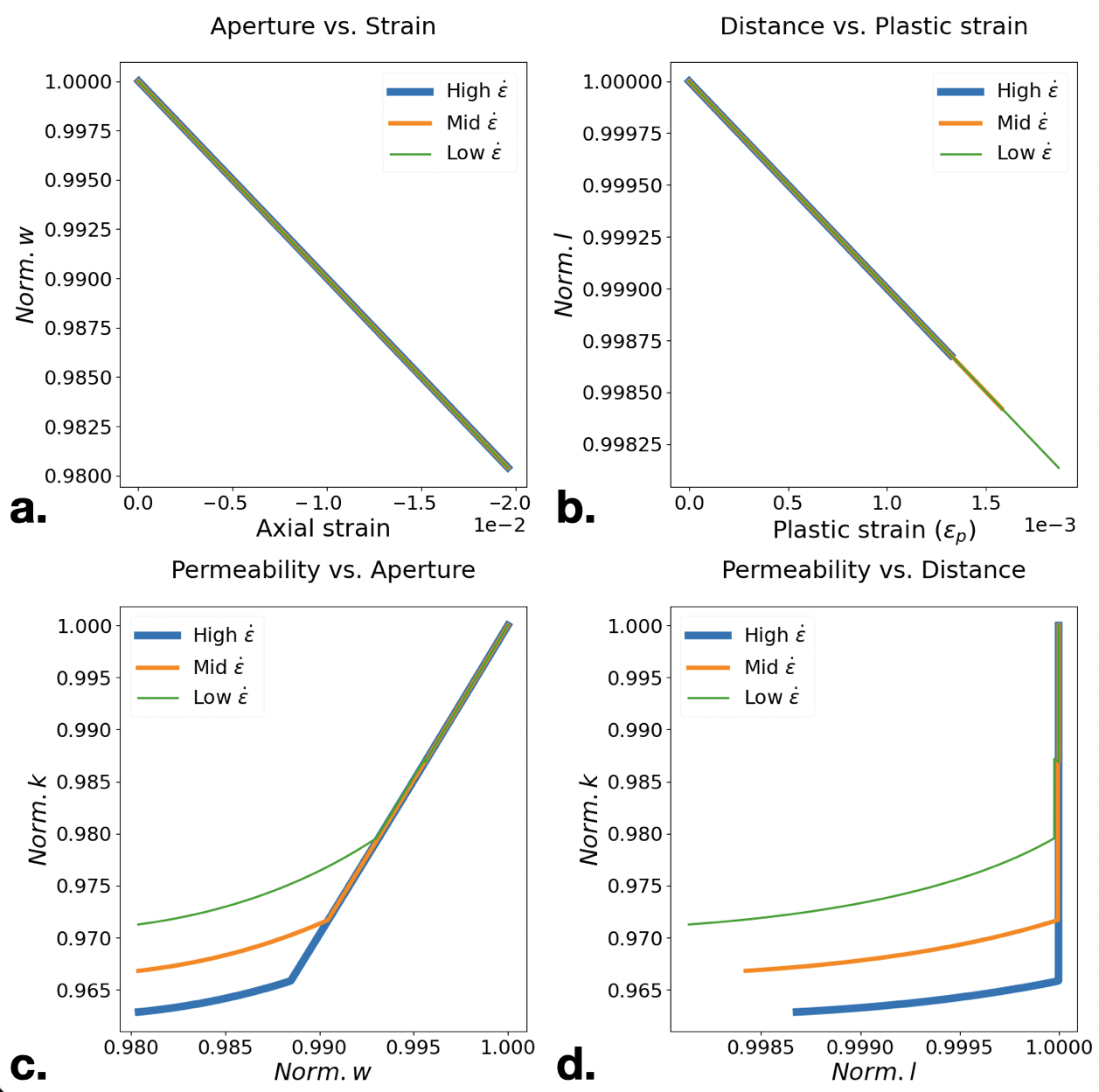}
\caption{Stress to Crack Geometry to Permeability mapping under Triaxial Compression for different strain rates: a. aperture vs axial strain (negative sign for compressive), b. distance vs plastic strain, c. permeability vs aperture and d. permeability vs distance. All crack geometries and permeability are normalized with respect to their initial values. Crack length results do not vary.} \label{simple_triaxial}
\end{figure}

Figure~\ref{simple_triaxial}a shows that aperture decreases with increasing compressive strain. This behavior follows the crack closure relations in Eqs.~\ref{elastic_opening} and \ref{wstrain}. Because crack length does not increase, aperture closure is the dominant mechanism that reduces permeability. Notice that in this case there is no strain rate dependency on aperture closure.

Figure~\ref{simple_triaxial}b shows the evolution of crack distance as a function of plastic strain. Plastic strain accumulates under compression and reduces the representative crack distance according to Eq.~\ref{ldens}. Lower strain rates induce larger final plastic strains as formulated in the material model, reducing crack distances. 

Figures~\ref{simple_triaxial}c and \ref{simple_triaxial}d show the competition between aperture closure, which lowers permeability, and distance reduction, which increases permeability. Inspecting Figure~\ref{simple_triaxial}c from right to left: permeability first decreases as aperture closes under compression, this effect is then lessened once plastic strain is attained and crack distance reduced. Notice that permeability keeps decreasing, suggesting that crack closure is the dominant mechanism. Strain rate effects are also noticeable in the early generation of plastic strain for low strain rates, which prompts distance reduction and ultimately affects the final permeability.

\subsection{Multiple-elements uniaxial setup and results}

\begin{figure}[ht!]  
\centering 
\noindent\includegraphics[width=\textwidth]{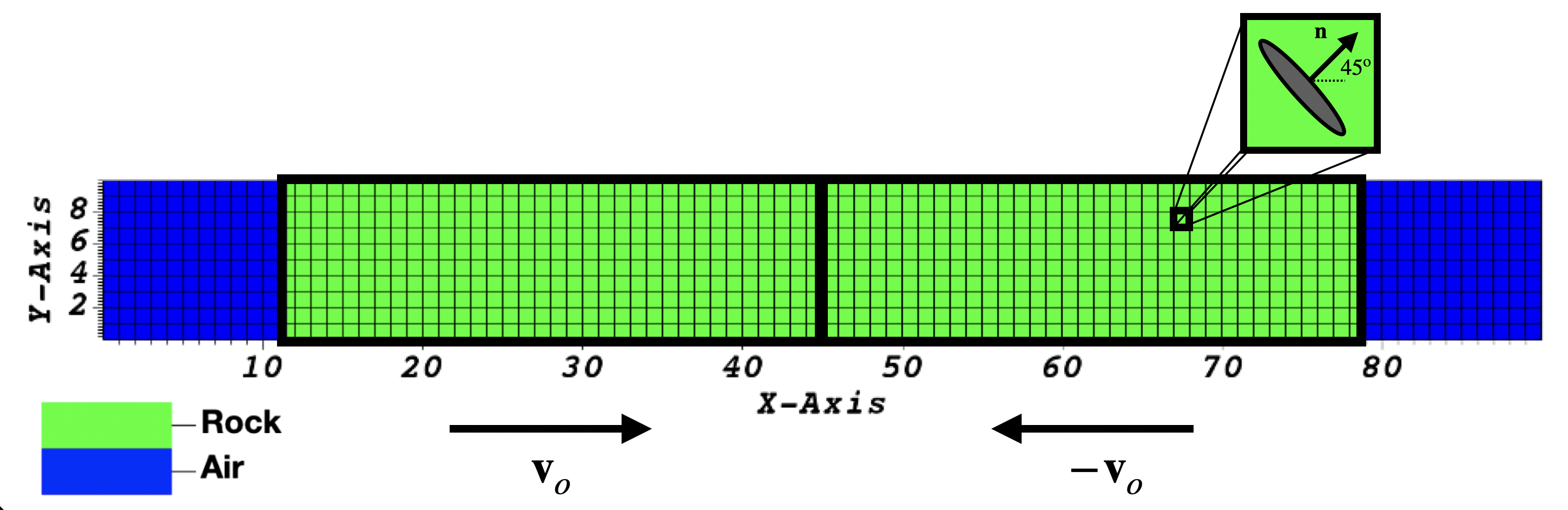}
\caption{Setup conditions for symmetric rock impact and orientation of crack normal.}  \label{symgeo}
\end{figure}

In this subsection we extend the single-element analysis to a multiple-element configuration, and we use an impact test problem to examine permeability evolution. The simulation consists of a symmetric impact in the $x$-direction between a pair of two-dimensional granite bars with identical material properties (Table~\ref{simpleT}) as shown in Figure~\ref{symgeo}. The computational domain contains $90 \times 10$ elements; each bar is $34~\mathrm{mm}$ long and $10~\mathrm{mm}$ wide. After impact, a non-separation condition at the contact surface is applied and the two bars behave as a single object. Symmetric boundary conditions are imposed in the $y$-direction to approximate a uniaxial strain problem, and outflow boundary conditions are applied in the $x$-direction. Moreover, to observe the individual effects of each variable more clearly, we have decoupled crack opening and crack length by substituting $\bar{c}/A$ with $\bar{w}^o$ in Eqs.~\ref{elastic_opening_ar} and \ref{w_general}. The bars are assigned equal and opposite initial velocities of $\mathbf{v}_{0}=(\pm 0.9, 0,0)~\mathrm{m\,s^{-1}}$. We use the GeoDyn finite volume Godunov Eulerian massively parallel code \citep{osti_928156,Vitali2012}, including its constitutive material library \citep{Vorobiev2021}. The orientation of the crack normal vector, specified in Table~\ref{simpleT} and shown in Figure~\ref{symgeo}, is prescribed so that both Mode~I and Mode~II crack propagation can develop. For conciseness, we focus on the permeability evolution in the horizontal $x$-direction; therefore, we report only the $k_{xx}$ component. Although our formulation has the capability to represent multiple crack orientations and permeability tensor components, this uniaxial configuration facilitates comparison with the single-element tests presented in the previous section.

\begin{figure}[ht!]  
\centering 
\noindent\includegraphics[width=\textwidth]{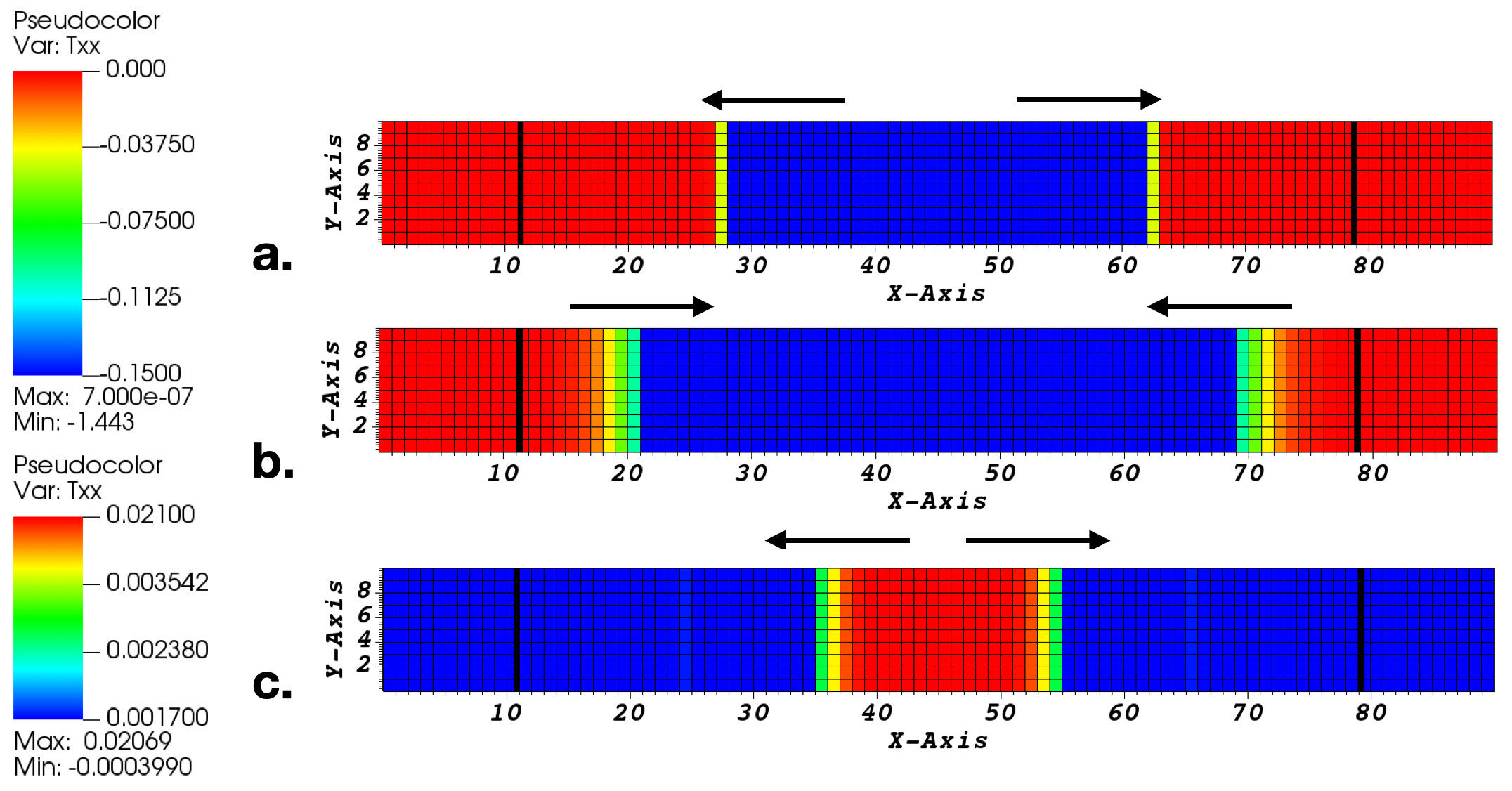}
\caption{Stress (GPa) in the $x$-direction, $\sigma_{xx}$, for symmetric rock impact at: a. onset of the compressive wave due to impact, expanding outward; b. development of rarefaction at the free ends, returning inward; and c. tensile stresses propagate outward, after rarefaction waves return to the impact region.}   \label{symT}
\end{figure}

Figure~\ref{symT} shows the axial stress component $\sigma_{xx}$ at three stages occurring during the impact. First, Figure~\ref{symT}a shows how the collision generates compressive waves that propagate from the impact point at the center of the domain toward the free ends of the bars. Then, following Figure~\ref{symT}b, the compressive waves reflect from the free ends towards the center as rarefaction waves. Finally in Figure~\ref{symT}c we observe that, after the rarefaction waves converge back to the impact point, a tensile stress regime develops due to the non-separation condition. This tension propagates outward and we display it using a different colorbar for Figure~\ref{symT}c. These three stages provide the reference sequence used next to interpret changes in crack aperture, crack length, crack distance, matrix porosity, and the corresponding permeability evolution.

\begin{figure}[ht!]  
\centering 
\noindent\includegraphics[width=\textwidth]{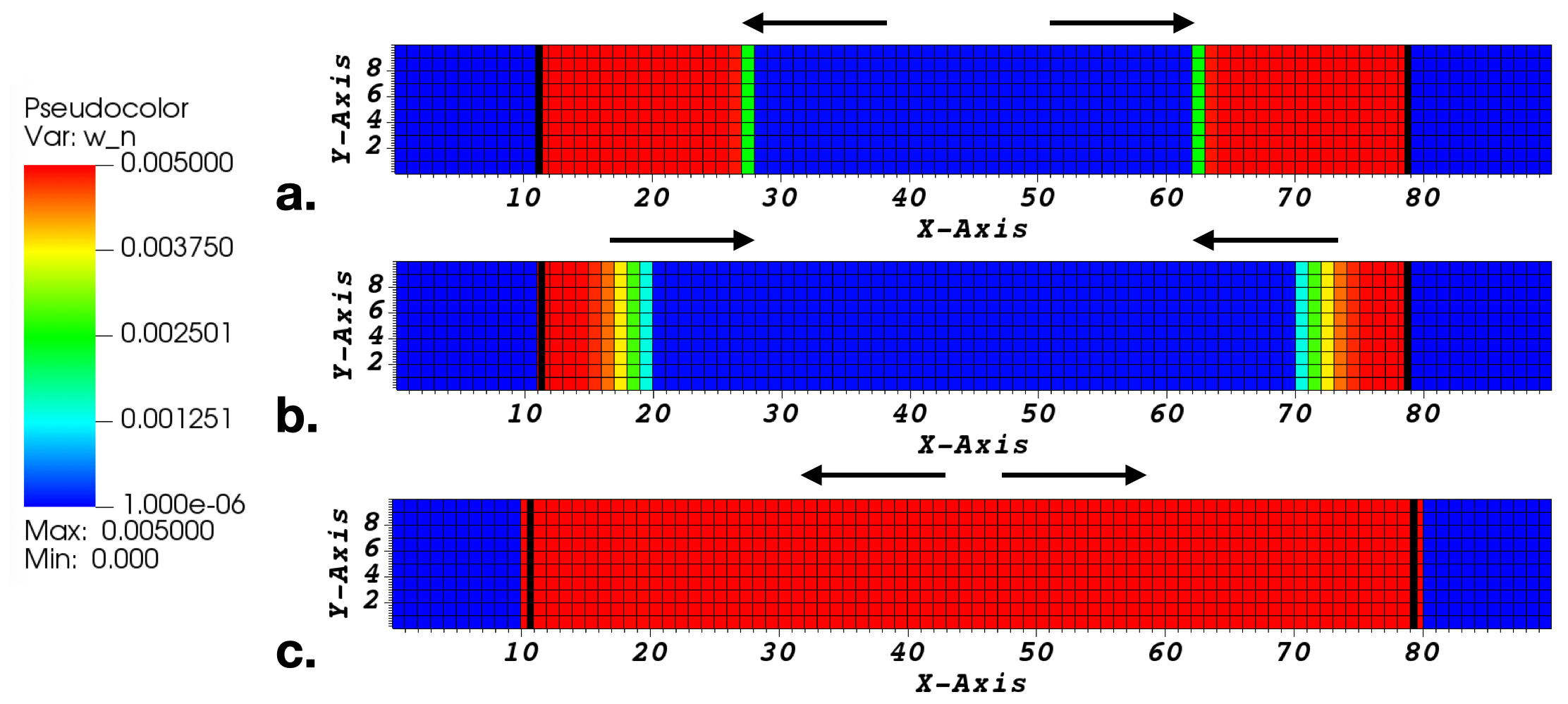}
\caption{Crack aperture in mm, in the direction of the normal vector $\mathbf{n}$, for symmetric rock impact at: a. onset of the compressive wave due to impact, b. development of rarefaction at the free ends, and c. tensile stresses propagate outward, after rarefaction waves return to the impact region.}   
\label{symw}
\end{figure}

Figure~\ref{symw} shows the response of crack aperture to the loading conditions. In Figure~\ref{symw}a we see how, during the compressive stage, aperture decreases to its minimal value $\bar{w}^\textrm{min}$ as cracks close elastically. This behavior is consistent with Eq.~\ref{elastic_opening} and Table \ref{simpleT}. As rarefaction waves propagate from the free ends of the bars, aperture increases towards its initial value (Figure~\ref{symw}b). Figure~\ref{symw}c displays the tensile regime where no appreciable change in aperture is noticeable because material failure relieves the crack normal stress and eliminates the shear stress. 

\begin{figure}[ht!]  
\centering 
\noindent\includegraphics[width=\textwidth]{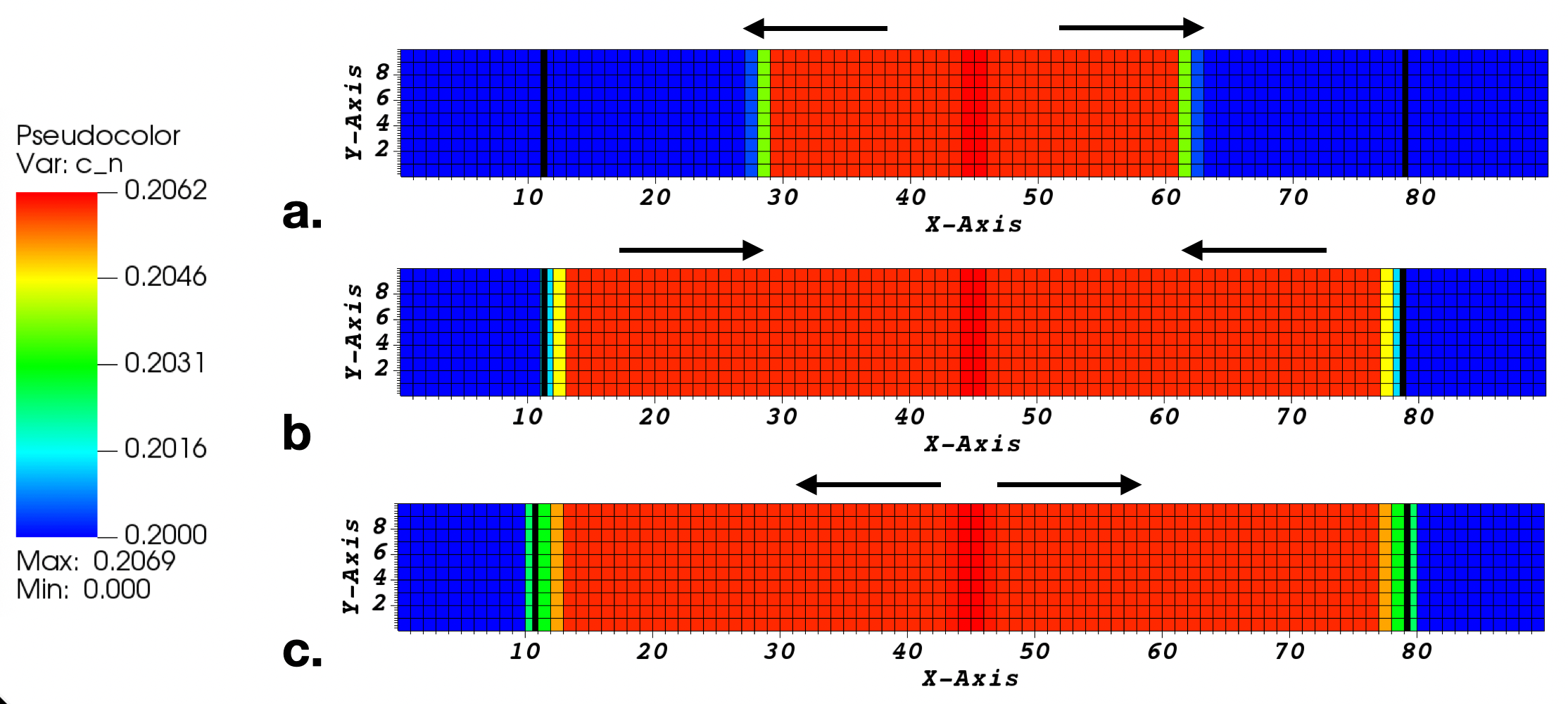}
\caption{Crack length in mm, in the direction of the normal vector $\mathbf{n}$, for symmetric rock impact at: a. onset of the compressive wave due to impact, b. development of rarefaction at the free ends, and c. tensile stresses propagate outward, after rarefaction waves return to the impact region.}   
\label{symc}
\end{figure}

Crack length evolution is described in Figure~\ref{symc}. Figure~\ref{symc}a shows how crack length grows during the compressive regime as waves move towards the ends of the bars. This growth is associated with Mode~II fracture, described in the compressive regime by Eq.~\textit{SM}-21. Figure~\ref{symc}b displays the subsequent rarefaction, which does not produce any crack growth since stresses return to zero (Figure~\ref{symT}b). Finally, in Figure \ref{symc}c, we observe that the tensile regime does not produce additional crack growth. This is a consequence of material failure, which eliminates the shear stress and reduces the normal stress, making the energy release rate in tension (Eq.~\textit{SM}-19) lower than the one experienced in the compressive regime.

\begin{figure}[ht!]  
\centering 
\noindent\includegraphics[width=\textwidth]{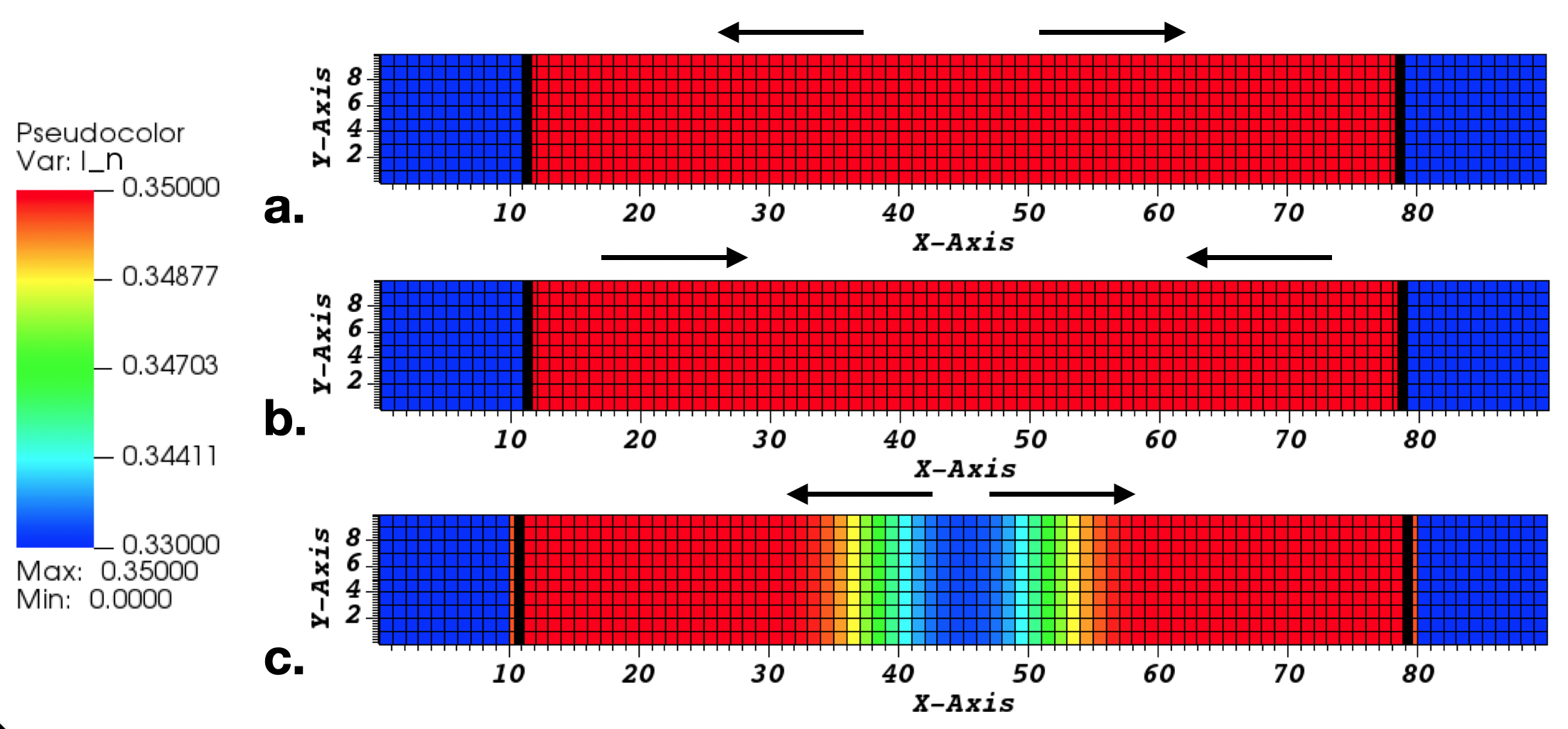}
\caption{Crack distance in mm, in the direction of the normal vector $\mathbf{n}$, for symmetric rock impact at: a. onset of the compressive wave due to impact, b. development of rarefaction at the free ends, and c. tensile stresses propagate outward, after rarefaction waves return to the impact region.} 
\label{syml}
\end{figure}

Figure~\ref{syml} presents the evolution of crack distance. Figure~\ref{syml}a and Figure \ref{syml}b display how crack distance is not modified during the compressive and rarefaction stages due to the absence of plastic strains. Figure~\ref{syml}c displays reduction in crack distance during the tensile regime because of material failure \citep{Vorobiev2021}, which develops plastic strain. This response follows Eq.~\ref{ldens} and represents nucleation-driven crack densification.

\begin{figure}[ht!]  
\centering 
\noindent\includegraphics[width=\textwidth]{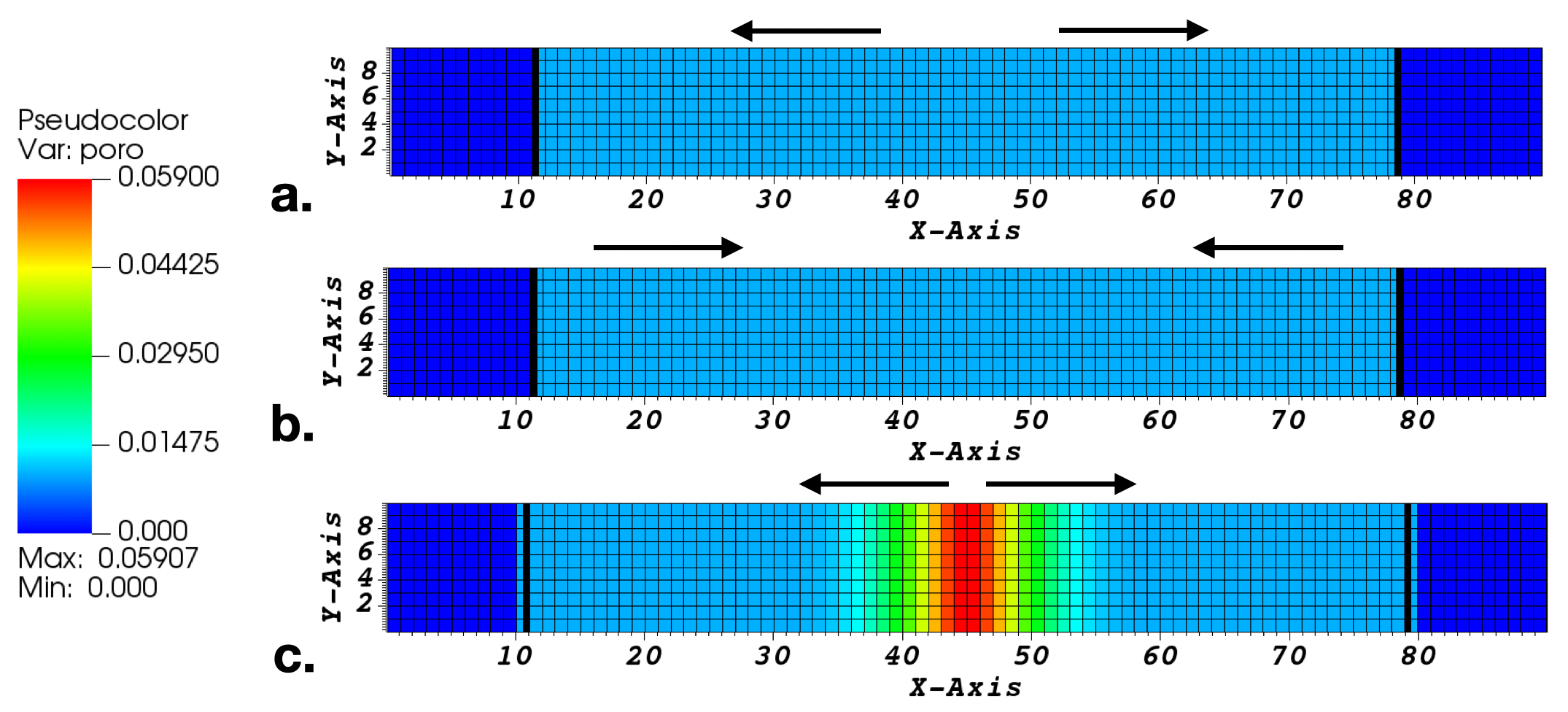}
\caption{Matrix porosity for symmetric rock impact at: a. onset of the compressive wave due to impact, b. development of rarefaction at the free ends, and c. tensile stresses propagate outward, after rarefaction waves return to the impact region.}    
\label{symporo}
\end{figure}

Matrix porosity evolution is shown in Figure~\ref{symporo}.  The compressive stress displayed in Figure~\ref{symporo}a induces negligible poroelastic effects, leaving the porosity essentially unchanged. Porosity also remains unchanged during the rarefaction regime in Figure~\ref{symporo}b as stresses return to zero. However, porosity increases during the tensile regime (Figure~\ref{symporo}c) due to material failure and dilatancy \citep{Vorobiev2021}. 

\begin{figure}[ht!]  
\centering 
\noindent\includegraphics[width=\textwidth]{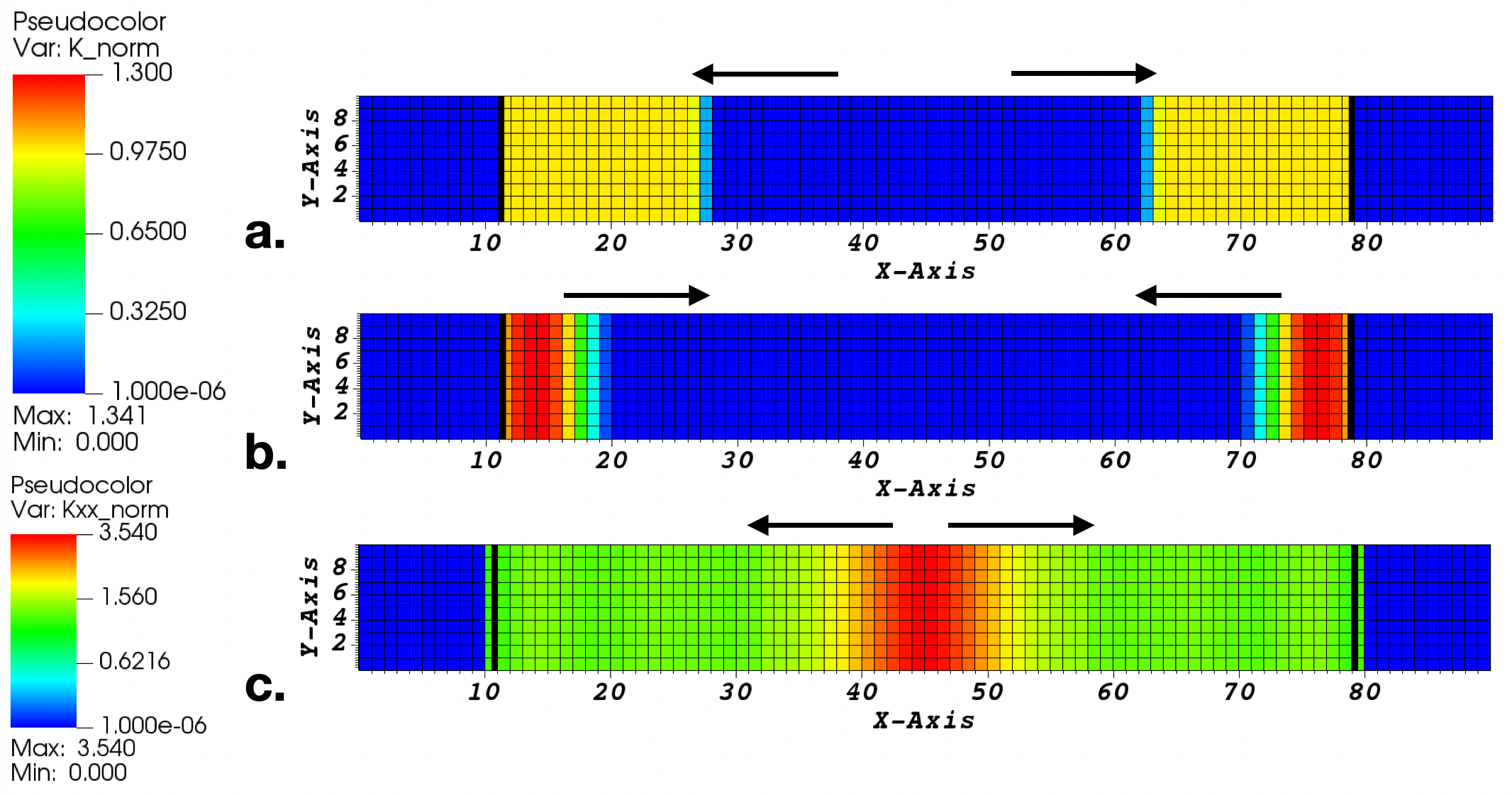}
\caption{Normalized permeability, with respect to its initial value, in the $x$-direction, $k_{xx}$, for symmetric rock impact at: a. onset of the compressive wave due to impact, b. development of rarefaction at the free ends, and c. tensile stresses propagate outward, after rarefaction waves return to the impact region.}    
\label{symK}
\end{figure}

Figure~\ref{symK} shows the evolution of the normalized permeability component $k_{xx}$ with respect to the initial permeability. During the initial compressive stage, displayed in Figure~\ref{symK}a, permeability decreases due to crack aperture reduction (see Figure~\ref{symw}a) while it increases due to crack length growth (Figure~\ref{symc}a). However, crack aperture is the dominant mechanism because it is reduced to its minimal value (Eq.~\ref{elastic_opening}) while crack length experiences a modest increase (see Eq.~\ref{include_eq}). In Figure~\ref{symK}b, as compression relaxes during rarefaction and the original aperture is reinstated (see Figure~\ref{symw}b), we observe a permeability increase due to the irreversible crack length growth experienced during compression (Figures~\ref{symc}a and b). Finally, Figure~\ref{symK}c shows the tensile regime where the material experiences a large permeability increment caused by the decrease in crack distance (see Figure~\ref{syml}c) and increase in matrix porosity (see Figure~\ref{symporo}c). We modify the colorbar scheme for this last regime for better visualization of the permeability rise. 

\section{Discussion}

The previous section showed the model's consistency with fundamental physical principles where permeability in low-porosity brittle rocks is represented by crack geometric microvariables.

In the single-element tests, different loading paths activate distinct combinations of crack aperture, crack length, crack distance, and connectivity. In the tensile and shear cases, crack propagation and aperture increase while enhancing fracture connectivity. Consequently, permeability increases by several orders of magnitude relative to its initial value. Strain rate effects on length and aperture are also noticeable in both cases. In the compression case, however, permeability is reduced because the elastic crack closure dominates permeability over crack distance reduction. In this case strain rate effects are present on distance. These tests demonstrate that even in a simplified setting, the proposed formulation captures the dependency of permeability evolution on loading conditions and strain rate.

In the multiple-element impact test, the results demonstrate the improved capability of the current permeability evolution model compared to a matrix porosity-based model. In the compression and rarefaction regimes our model predicts permeability changes while matrix porosity is preserved. The effects of matrix porosity on permeability are noticeable only in the tensile regime. These results shows that our crack-based formulation can capture permeability changes that would be ignored if only matrix porosity is considered.

Our results are consistent with key experimental and in situ trends reported for brittle rocks, including permeability reduction during crack closure under compression, sharp permeability increase during fracture growth, and sensitivity to crack connectivity \citep{Zhang2022}. The model also reproduces order-of-magnitude permeability increases under dynamic loading, consistent with observations such as those reported by \cite{Aben2020}. While these results are encouraging, being able to correlate permeability with each microcrack variable at different loading conditions would be ideal for quantitative validation. Initial and post-mortem data could be obtained from controlled experiments using microscopy or X-ray computed tomography on three-dimensional printed samples with prescribed crack geometries. Unfortunately, obtaining transitional states from laboratory experiments seems unfeasible. However, numerical models with discrete fracture networks could be a viable solution to generate these correlations despite model uncertainty.

The model discussed in this paper demonstrates a significant improvement over classic matrix porosity-based formulation, but it could be further enhanced. Accounting for crack rotation will enable the model to represent problems with substantial rotational deformation. Introducing multiple cracks families within each element would improve the accuracy of crack network representation. Generalizing the crack coalescence formulation to account for different crack orientations will also result in a more accurate crack network representation. Multi-dimensional crack nucleation will improve crack distance accuracy. The addition of roughness to the crack surface will introduce the concept of hydraulic aperture, improving permeability predictions. Accounting for ductile crack propagation will extend the range of applicable materials. The transition of microcracks into discrete cracks would help bridge the microscale and macroscale description of fracture evolution. Finally, the model could benefit from a different microstructural idealized geometry such as cylinders, which are a better representation for high-porosity rocks (e.g., porous volcanic rocks). Extending the present microvariable framework to such improvements would broaden the model applicability while preserving its central advantage: permeability evolution that is physically interpretable from pore scale variables. 

This work provides the basis for predicting both permeability reduction and enhancement in dynamically loaded low-porosity rocks. The framework accounts for spatial stress variability and its effects on crack aperture, propagation, nucleation, and connectivity. Despite its infancy, the model demonstrates that permeability can be tracked in a physically meaningful and computationally tractable manner by evolving a reduced set of crack geometric microvariables.  

\section{Conclusions}
We developed a penny-shaped crack-based permeability evolution model for low-porosity brittle rocks in which microcrack networks control the flow. The formulation links permeability to a reduced set of evolving crack geometric microvariables: aperture, length, and distance. Through these variables, the model represents porosity, conductance, and connectivity in a physically interpretable manner. This formulation provides an alternative to permeability evolution models based primarily on porosity, effective pressure, or empirical calibration.

The proposed model incorporates several physics-based mechanisms. Crack aperture can open and close through elastic and inelastic deformation, resulting in reversible permeability changes. The evolution of crack propagation, nucleation, and coalescence is irreversible and induces permanent permeability increase. Any of these mechanisms can dominate permeability based on initial conditions, loading type (e.g., simple shear, tension, compression, etc.), strain rate, and crack orientation. 

This evolution model can incorporate the matrix porosity predicted by a poromechanical model into the total porosity used to calculate permeability. The addition of matrix porosity enables the model to combine the classical Kozeny--Carman porosity-driven permeability with the crack network-based permeability. Notice that the model requires constitutive description of the equation of state, material strength and failure response.

Overall, this work provides a continuum framework for linking dynamic mechanical loading, microcrack geometry, and permeability evolution for low-porosity brittle rocks. This approach improves classical permeability predictions in rocks by capturing the evolution of crack aperture, length and distance. Future work should extend the current formulation to accommodate tensorial permeability, evolving crack orientations, multiple crack families, crack roughness, alternative crack geometries, macro fractures, and hydraulic aperture effects. These additions will enable our model to represent more complex environments. 

\section{Data availability}

\noindent All data supporting the findings of this study are publicly available. The complete set of input files used to initialize the numerical simulations, along with the corresponding output datasets required to reproduce all figures and quantitative results presented in this manuscript, are archived at Zenodo: \url{https://zenodo.org/records/19474991}.

\noindent Simulations were performed using the in-house finite element code GeoDyn. Its numerical implementation, solution algorithms, and poromechanical material library are documented in \cite{osti_928156, Vitali2012, Vorobiev2021}. The constitutive formulations used in this study are fully described in those references and are sufficient to enable independent verification and reproduction of the reported results.

\section{CRediT authorship contribution statement}

Rigoberto Moncada: Conceptualization, Methodology, Formal analysis, Investigation, Software, Validation, Visualization, Writing -- original draft, Writing -- review and editing.

Efrem Vitali: Conceptualization, Methodology, Supervision, Project administration, Resources, Writing -- review and editing.

\section{Declaration of competing interest}
The authors declare that they have no known competing financial interests or personal relationships that could have appeared to influence the work reported in this paper.

\section{Funding}
This Low Yield Nuclear Monitoring (LYNM) research was funded by the National Nuclear Security Administration, Defense Nuclear Nonproliferation Research and Development (NNSA DNN R\&D).

\section{Acknowledgements}
The authors acknowledge important interdisciplinary collaboration with scientists and engineers from LANL, LLNL, NNSS, PNNL, and SNL. Research was developed and the manuscript was written under the auspices of the U.S. Department of Energy by Lawrence Livermore National Laboratory under Contract DE-AC52-07NA27344.

\bibliographystyle{elsarticle-harv}
\bibliography{references}

@article{Benson2006,
	title = {Role of void space geometry in permeability evolution in crustal rocks at elevated pressure},
	volume = {111},
	ISSN = {0148-0227},
	url = {http://dx.doi.org/10.1029/2006JB004309},
	DOI = {10.1029/2006jb004309},
	number = {B12},
	journal = {Journal of Geophysical Research: Solid Earth},
	publisher = {American Geophysical Union (AGU)},
	author = {Benson,  Philip M. and Meredith,  Philip G. and Schubnel,  Alexandre},
	year = {2006},
	month = dec 
}

@article{Gueguen1989,
	title = {Transport properties of rocks from statistics and percolation},
	volume = {21},
	ISSN = {1573-8868},
	url = {http://dx.doi.org/10.1007/BF00897237},
	DOI = {10.1007/bf00897237},
	number = {1},
	journal = {Mathematical Geology},
	publisher = {Springer Science and Business Media LLC},
	author = {Gueguen,  Y. and Dienes,  J.},
	year = {1989},
	month = jan,
	pages = {1–13}
}

@article{Peach1996,
	title = {Influence of crystal plastic deformation on dilatancy and permeability development in synthetic salt rock},
	volume = {256},
	ISSN = {0040-1951},
	url = {http://dx.doi.org/10.1016/0040-1951(95)00170-0},
	DOI = {10.1016/0040-1951(95)00170-0},
	number = {1–4},
	journal = {Tectonophysics},
	publisher = {Elsevier BV},
	author = {Peach,  Colin J. and Spiers,  Christopher J.},
	year = {1996},
	month = may,
	pages = {101–128}
}

@article{Zhang2023,
	title = {Study on permeability evolution and damage mechanism along the EGS fracture in heat mining stage under thermal stress/cracking},
	volume = {11},
	ISSN = {2195-9706},
	url = {http://dx.doi.org/10.1186/s40517-023-00274-2},
	DOI = {10.1186/s40517-023-00274-2},
	number = {1},
	journal = {Geothermal Energy},
	publisher = {Springer Science and Business Media LLC},
	author = {Zhang,  Wei and Wang,  Dong and Wang,  Zenglin and Guo,  Tiankui and Wang,  Chunguang and He,  Jiayuan and Zhang,  Le and Zheng,  Peng and Qu,  Zhanqing},
	year = {2023},
	month = nov 
}

@book{bear1972,
	title     = "Dynamics of Fluids in Porous Media",
	author    = "Bear, Jacob",
	year      = 1972,
	publisher = "Elsevier",
	address   = "N.Y."
}

@article{Aben2020,
  title = {Variation of Hydraulic Properties Due to Dynamic Fracture Damage: Implications for Fault Zones},
  volume = {125},
  ISSN = {2169-9356},
  url = {http://dx.doi.org/10.1029/2019JB018919},
  DOI = {10.1029/2019jb018919},
  number = {4},
  journal = {Journal of Geophysical Research: Solid Earth},
  publisher = {American Geophysical Union (AGU)},
  author = {Aben,  Franciscus M. and Doan,  Mai‐Linh and Mitchell,  Thomas M.},
  year = {2020},
  month = apr 
}

@article{Bourret2019,
	title = {Evaluating the Importance of Barometric Pumping for Subsurface Gas Transport Near an Underground Nuclear Test Site},
	volume = {18},
	ISSN = {1539-1663},
	url = {http://dx.doi.org/10.2136/vzj2018.07.0134},
	DOI = {10.2136/vzj2018.07.0134},
	number = {1},
	journal = {Vadose Zone Journal},
	publisher = {Wiley},
	author = {Bourret,  S.M. and Kwicklis,  E.M. and Miller,  T.A. and Stauffer,  P.H.},
	year = {2019},
	month = jan,
	pages = {1–17}
}

@article{DazCuriel2022,
	title = {New granulometric expressions for estimating permeability of granular drainages},
	volume = {81},
	ISSN = {1435-9537},
	url = {http://dx.doi.org/10.1007/s10064-022-02897-4},
	DOI = {10.1007/s10064-022-02897-4},
	number = {10},
	journal = {Bulletin of Engineering Geology and the Environment},
	publisher = {Springer Science and Business Media LLC},
	author = {Díaz-Curiel,  Jesús and Miguel,  María J. and Biosca,  Bárbara and Arévalo-Lomas,  Lucía},
	year = {2022},
	month = sep 
}

@article{Deng2022,
  title = {Study on crack evolutional behavior of rocks in triaxial compression based on colony growth dynamics model},
  volume = {12},
  ISSN = {2045-2322},
  url = {http://dx.doi.org/10.1038/s41598-022-23177-x},
  DOI = {10.1038/s41598-022-23177-x},
  number = {1},
  journal = {Scientific Reports},
  publisher = {Springer Science and Business Media LLC},
  author = {Deng,  Naifu and Qiao,  Lan and Li,  Qingwen and Hao,  Jiawang and Wei,  Mengxi and Zhang,  Qinglong},
  year = {2022},
  month = nov 
}

@inproceedings{dienes1978statistical,
	title={A statistical theory of fragmentation},
	author={Dienes, John K},
	booktitle={ARMA US Rock Mechanics/Geomechanics Symposium},
	pages={ARMA--78},
	year={1978},
	organization={ARMA}
}

@article{Dienes2006,
	title = {Impact initiation of explosives and propellants via statistical crack mechanics},
	volume = {54},
	ISSN = {0022-5096},
	url = {http://dx.doi.org/10.1016/j.jmps.2005.12.001},
	DOI = {10.1016/j.jmps.2005.12.001},
	number = {6},
	journal = {Journal of the Mechanics and Physics of Solids},
	publisher = {Elsevier BV},
	author = {Dienes,  J.K. and Zuo,  Q.H. and Kershner,  J.D.},
	year = {2006},
	month = jun,
	pages = {1237–1275}
}

@article{Fuchs2025,
  title = {Permeability Evolution and Gouge Formation During Fracture Shearing},
  volume = {52},
  ISSN = {1944-8007},
  url = {http://dx.doi.org/10.1029/2025GL117217},
  DOI = {10.1029/2025gl117217},
  number = {23},
  journal = {Geophysical Research Letters},
  publisher = {American Geophysical Union (AGU)},
  author = {Fuchs,  Marco and Blum,  Philipp and Bl\"{o}cher,  Guido and Scholtès,  Luc},
  year = {2025},
  month = dec 
}

@article{Gao2022,
	title = {A critical review of coal permeability models},
	volume = {326},
	ISSN = {0016-2361},
	url = {http://dx.doi.org/10.1016/j.fuel.2022.125124},
	DOI = {10.1016/j.fuel.2022.125124},
	journal = {Fuel},
	publisher = {Elsevier BV},
	author = {Gao,  Qi and Liu,  Jishan and Huang,  Yifan and Li,  Wai and Shi,  Rui and Leong,  Yee-Kwong and Elsworth,  Derek},
	year = {2022},
	month = oct,
	pages = {125124}
}

@article{Gavrilenko1989,
	title = {Pressure dependence of permeability: a model for cracked rocks},
	volume = {98},
	ISSN = {1365-246X},
	url = {http://dx.doi.org/10.1111/j.1365-246X.1989.tb05521.x},
	DOI = {10.1111/j.1365-246x.1989.tb05521.x},
	number = {1},
	journal = {Geophysical Journal International},
	publisher = {Oxford University Press (OUP)},
	author = {Gavrilenko,  P. and Gueguen,  Y.},
	year = {1989},
	month = jul,
	pages = {159–172}
}

@article{Gehne2019,
  title = {Permeability enhancement through hydraulic fracturing: laboratory measurements combining a 3D printed jacket and pore fluid over-pressure},
  volume = {9},
  ISSN = {2045-2322},
  url = {http://dx.doi.org/10.1038/s41598-019-49093-1},
  DOI = {10.1038/s41598-019-49093-1},
  number = {1},
  journal = {Scientific Reports},
  publisher = {Springer Science and Business Media LLC},
  author = {Gehne,  Stephan and Benson,  Philip M.},
  year = {2019},
  month = aug 
}

@article{Heath2021,
  title = {Heterogeneous multiphase flow properties of volcanic rocks and implications for noble gas transport from underground nuclear explosions},
  volume = {20},
  ISSN = {1539-1663},
  url = {http://dx.doi.org/10.1002/vzj2.20123},
  DOI = {10.1002/vzj2.20123},
  number = {3},
  journal = {Vadose Zone Journal},
  publisher = {Wiley},
  author = {Heath,  Jason E. and Kuhlman,  Kristopher L. and Broome,  Scott T. and Wilson,  Jennifer E. and Malama,  Bwalya},
  year = {2021},
  month = may 
}

@article{Hosseinzadegan2023,
  title = {Review on pore-network modeling studies of gas-condensate flow: Pore structure,  mechanisms,  and implementations},
  volume = {226},
  ISSN = {2949-8910},
  url = {http://dx.doi.org/10.1016/j.geoen.2023.211693},
  DOI = {10.1016/j.geoen.2023.211693},
  journal = {Geoenergy Science and Engineering},
  publisher = {Elsevier BV},
  author = {Hosseinzadegan,  Ahmad and Raoof,  Amir and Mahdiyar,  Hojjat and Nikooee,  Ehsan and Ghaedi,  Mojtaba and Qajar,  Jafar},
  year = {2023},
  month = jul,
  pages = {211693}
}

@inproceedings{inproceedingsLi2024,
author = {Bin, Li and Li, Qi and Shen, Haimeng},
year = {2024},
month = {11},
pages = {},
title = {Numerical simulation of the dynamic evolution of permeability during the rough fractured rock shear process under different confining pressures}
}

@article{Li2025,
  title = {Research progress and scientific challenges in permeability evolution of hydrate bearing sediments},
  ISSN = {1995-8226},
  url = {http://dx.doi.org/10.1016/j.petsci.2025.09.007},
  DOI = {10.1016/j.petsci.2025.09.007},
  journal = {Petroleum Science},
  publisher = {Elsevier BV},
  author = {Li,  Yao-Bin and Xin,  Xin and Zhu,  Hui-Xing and Su,  Yue and Yuan,  Yi-Long and Xu,  Tian-Fu},
  year = {2025},
  month = sep 
}

@article{Liao2023,
	title = {An anisotropic damage–permeability model for hydraulic fracturing in hard rock},
	volume = {18},
	ISSN = {1861-1133},
	url = {http://dx.doi.org/10.1007/s11440-022-01793-1},
	DOI = {10.1007/s11440-022-01793-1},
	number = {7},
	journal = {Acta Geotechnica},
	publisher = {Springer Science and Business Media LLC},
	author = {Liao,  Jianxing and Wang,  Hong and Mehmood,  Faisal and Cheng,  Cao and Hou,  Zhengmeng},
	year = {2023},
	month = feb,
	pages = {3661–3681}
}

@article{Lubarda1993,
	title = {Damage tensors and the crack density distribution},
	volume = {30},
	ISSN = {0020-7683},
	url = {http://dx.doi.org/10.1016/0020-7683(93)90158-4},
	DOI = {10.1016/0020-7683(93)90158-4},
	number = {20},
	journal = {International Journal of Solids and Structures},
	publisher = {Elsevier BV},
	author = {Lubarda,  V.A. and Krajcinovic,  D.},
	year = {1993},
	pages = {2859–2877}
}

@article{Luo2023,
  title = {Estimation of 3D Permeability from Pore Network Models Constructed Using 2D Thin-Section Images in Sandstone Reservoirs},
  volume = {16},
  ISSN = {1996-1073},
  url = {http://dx.doi.org/10.3390/en16196976},
  DOI = {10.3390/en16196976},
  number = {19},
  journal = {Energies},
  publisher = {MDPI AG},
  author = {Luo,  Chengfei and Wan,  Huan and Chen,  Jinding and Huang,  Xiangsheng and Cui,  Shuheng and Qin,  Jungan and Yan,  Zhuoyu and Qiao,  Dan and Shi,  Zhiqiang},
  year = {2023},
  month = oct,
  pages = {6976}
}

@article{Ma2015,
	title = {Review of permeability evolution model for fractured porous media},
	volume = {7},
	ISSN = {1674-7755},
	url = {http://dx.doi.org/10.1016/j.jrmge.2014.12.003},
	DOI = {10.1016/j.jrmge.2014.12.003},
	number = {3},
	journal = {Journal of Rock Mechanics and Geotechnical Engineering},
	publisher = {Elsevier BV},
	author = {Ma,  Jianjun},
	year = {2015},
	month = jun,
	pages = {351–357}
}

@article{Mayrhofer2019,
  title = {Universal and Nonuniversal Aperture‐to‐Length Scaling of Opening Mode Fractures Developing in a Particle‐Based Lattice Solid Model},
  volume = {124},
  ISSN = {2169-9356},
  url = {http://dx.doi.org/10.1029/2018JB015960},
  DOI = {10.1029/2018jb015960},
  number = {3},
  journal = {Journal of Geophysical Research: Solid Earth},
  publisher = {American Geophysical Union (AGU)},
  author = {Mayrhofer,  Franziska and Sch\"{o}pfer,  Martin P. J. and Grasemann,  Bernhard},
  year = {2019},
  month = mar,
  pages = {3197–3218}
}

@article{Meyer2024,
  title = {Permeability partitioning through the brittle-to-ductile transition and its implications for supercritical geothermal reservoirs},
  volume = {15},
  ISSN = {2041-1723},
  url = {http://dx.doi.org/10.1038/s41467-024-52092-0},
  DOI = {10.1038/s41467-024-52092-0},
  number = {1},
  journal = {Nature Communications},
  publisher = {Springer Science and Business Media LLC},
  author = {Meyer,  Gabriel G. and Shahin,  Ghassan and Cordonnier,  Benoît and Violay,  Marie},
  year = {2024},
  month = sep 
}

@article{Mitchell2008,
  title = {Experimental measurements of permeability evolution during triaxial compression of initially intact crystalline rocks and implications for fluid flow in fault zones},
  volume = {113},
  ISSN = {0148-0227},
  url = {http://dx.doi.org/10.1029/2008JB005588},
  DOI = {10.1029/2008jb005588},
  number = {B11},
  journal = {Journal of Geophysical Research: Solid Earth},
  publisher = {American Geophysical Union (AGU)},
  author = {Mitchell,  T. M. and Faulkner,  D. R.},
  year = {2008},
  month = nov 
}

@article{Mitchell2022,
  title = {Quantifying the Permeability Enhancement from Blast-Induced Microfractures in Porphyry Rocks Using a Cumulant Lattice Boltzmann Method},
  volume = {146},
  ISSN = {1573-1634},
  url = {http://dx.doi.org/10.1007/s11242-022-01875-4},
  DOI = {10.1007/s11242-022-01875-4},
  number = {3},
  journal = {Transport in Porous Media},
  publisher = {Springer Science and Business Media LLC},
  author = {Mitchell,  T. R. and Roslin,  A. and Laniewski-Wollk,  L. and Onederra,  I. and Leonardi,  C. R.},
  year = {2022},
  month = nov,
  pages = {587–615}
}

@article{Morris2003, 
	title={A constitutive model for stress‐induced permeability and porosity evolution of Berea sandstone}, 
	volume={108}, 
	ISSN={0148-0227}, 
	url={http://dx.doi.org/10.1029/2001JB000463}, 
	DOI={10.1029/2001jb000463}, 
	number={B10}, 
	journal={Journal of Geophysical Research: Solid Earth}, 
	publisher={American Geophysical Union (AGU)}, 
	author={Morris, J. P. and Lomov, I. N. and Glenn, L. A.}, 
	year={2003}, 
	month=oct 
}

@article{Morrow2001,
	title = {Permeability reduction in granite under hydrothermal conditions},
	volume = {106},
	ISSN = {0148-0227},
	url = {http://dx.doi.org/10.1029/2000JB000010},
	DOI = {10.1029/2000jb000010},
	number = {B12},
	journal = {Journal of Geophysical Research: Solid Earth},
	publisher = {American Geophysical Union (AGU)},
	author = {Morrow,  C. A. and Moore,  D. E. and Lockner,  D. A.},
	year = {2001},
	month = dec,
	pages = {30551–30560}
}

@article{Oda1985,
	title = {Permeability tensor for discontinuous rock masses},
	volume = {35},
	ISSN = {1751-7656},
	url = {http://dx.doi.org/10.1680/geot.1985.35.4.483},
	DOI = {10.1680/geot.1985.35.4.483},
	number = {4},
	journal = {Géotechnique},
	publisher = {Thomas Telford Ltd.},
	author = {Oda,  M.},
	year = {1985},
	month = dec,
	pages = {483–495}
}

@conference{osti_928156,
  author       = {Lomov, I and Pember, R and Greenough, J and Liu, B},
  title        = {Patch-based Adaptive Mesh Refinement for Multimaterial Hydrodynamics},
  url          = {https://www.osti.gov/biblio/928156},
  place        = {United States},
  organization = {Lawrence Livermore National Laboratory (LLNL), Livermore, CA},
  year         = {2005},
  month        = {10}}

@article{Pan2012,
	title = {Modelling permeability for coal reservoirs: A review of analytical models and testing data},
	volume = {92},
	ISSN = {0166-5162},
	url = {http://dx.doi.org/10.1016/j.coal.2011.12.009},
	DOI = {10.1016/j.coal.2011.12.009},
	journal = {International Journal of Coal Geology},
	publisher = {Elsevier BV},
	author = {Pan,  Zhejun and Connell,  Luke D.},
	year = {2012},
	month = mar,
	pages = {1–44}
}

@article{Paliwal2008, 
	title={An interacting micro-crack damage model for failure of brittle materials under compression}, 
	volume={56}, 
	ISSN={0022-5096}, url={http://dx.doi.org/10.1016/j.jmps.2007.06.012}, 
	DOI={10.1016/j.jmps.2007.06.012}, 
	number={3}, 
	journal={Journal of the Mechanics and Physics of Solids}, 
	publisher={Elsevier BV}, 
	author={Paliwal, B. and Ramesh, K.T.}, 
	year={2008}, 
	month=mar, 
	pages={896–923} 
}

@article{Perol2016, 
	title={Micromechanics-Based Permeability Evolution in Brittle Materials at High Strain Rates}, 
	volume={173}, ISSN={1420-9136}, 
	url={http://dx.doi.org/10.1007/s00024-016-1354-4}, 
	DOI={10.1007/s00024-016-1354-4}, 
	number={8}, journal={Pure and Applied Geophysics}, 
	publisher={Springer Science and Business Media LLC}, 
	author={Perol, Thibaut and Bhat, Harsha S.}, 
	year={2016}, 
	month=jul, 
	pages={2857–2868}
}

@article{Payton2022,
	title = {The upper percolation threshold and porosity–permeability relationship in sandstone reservoirs using digital image analysis},
	volume = {12},
	ISSN = {2045-2322},
	url = {http://dx.doi.org/10.1038/s41598-022-15651-3},
	DOI = {10.1038/s41598-022-15651-3},
	number = {1},
	journal = {Scientific Reports},
	publisher = {Springer Science and Business Media LLC},
	author = {Payton,  Ryan L. and Chiarella,  Domenico and Kingdon,  Andrew},
	year = {2022},
	month = jul 
}

@article{Peng2021, 
	title={A pore geometry-based permeability model for tight rocks and new sight of impact of stress on permeability}, 
	volume={91}, 
	ISSN={1875-5100}, 
	url={http://dx.doi.org/10.1016/j.jngse.2021.103958}, 
	DOI={10.1016/j.jngse.2021.103958}, journal={Journal of Natural Gas Science and Engineering}, 
	publisher={Elsevier BV}, 
	author={Peng, Yan and Liu, Jishan and Zhang, Guangqing and Pan, Zhejun and Ma, Zhixiao and Wang, Yibo and Hou, Yanan}, 
	year={2021}, 
	month=jul, 
	pages={103958}
}

@article{Pietruszczak2024,
  title = {Assessment of stress-induced evolution of permeability tensor in rock mass containing discrete fracture network},
  volume = {174},
  ISSN = {0266-352X},
  url = {http://dx.doi.org/10.1016/j.compgeo.2024.106580},
  DOI = {10.1016/j.compgeo.2024.106580},
  journal = {Computers and Geotechnics},
  publisher = {Elsevier BV},
  author = {Pietruszczak,  S. and Jameei,  A.A.},
  year = {2024},
  month = oct,
  pages = {106580}
}

@article{Roy2021,
  title = {From Nucleation to Percolation: The Effect of System Size when Disorder and Stress Localization Compete},
  volume = {9},
  ISSN = {2296-424X},
  url = {http://dx.doi.org/10.3389/fphy.2021.752086},
  DOI = {10.3389/fphy.2021.752086},
  journal = {Frontiers in Physics},
  publisher = {Frontiers Media SA},
  author = {Roy,  Subhadeep},
  year = {2021},
  month = nov 
}

@article{Simpson2001,
	title = {Permeability enhancement due to microcrack dilatancy in the damage regime},
	volume = {106},
	ISSN = {0148-0227},
	url = {http://dx.doi.org/10.1029/2000JB900194},
	DOI = {10.1029/2000jb900194},
	number = {B3},
	journal = {Journal of Geophysical Research: Solid Earth},
	publisher = {American Geophysical Union (AGU)},
	author = {Simpson,  Guy and Guéguen,  Yves and Schneider,  Frédéric},
	year = {2001},
	month = mar,
	pages = {3999–4016}
}

@article{Simpson2003,
	title = {Analytical Model for Permeability Evolution in Microcracking Rock},
	volume = {160},
	ISSN = {1420-9136},
	url = {http://dx.doi.org/10.1007/PL00012578},
	DOI = {10.1007/pl00012578},
	number = {5–6},
	journal = {Pure and Applied Geophysics},
	publisher = {Springer Science and Business Media LLC},
	author = {Simpson,  G. D. H. and Guéguen,  Y. and Schneider,  F.},
	year = {2003},
	month = may,
	pages = {999–1008}
}

@article{Sueyoshi2023,
  title = {Permeability evolution in fine‐grained Aji granite during triaxial compression experiments},
  volume = {72},
  ISSN = {1365-2478},
  url = {http://dx.doi.org/10.1111/1365-2478.13412},
  DOI = {10.1111/1365-2478.13412},
  number = {2},
  journal = {Geophysical Prospecting},
  publisher = {Wiley},
  author = {Sueyoshi,  Kazumasa and Katayama,  Ikuo and Sawayama,  Kazuki},
  year = {2023},
  month = aug,
  pages = {675–684}
}

@article{Thomas2020,
  title = {Permeability of Three‐Dimensional Numerically Grown Geomechanical Discrete Fracture Networks With Evolving Geometry and Mechanical Apertures},
  volume = {125},
  ISSN = {2169-9356},
  url = {http://dx.doi.org/10.1029/2019JB018899},
  DOI = {10.1029/2019jb018899},
  number = {4},
  journal = {Journal of Geophysical Research: Solid Earth},
  publisher = {American Geophysical Union (AGU)},
  author = {Thomas,  Robin N. and Paluszny,  Adriana and Zimmerman,  Robert W.},
  year = {2020},
  month = apr 
}

@inbook{Torquato_2002-ax, title={Some Continuum Percolation Results}, ISBN={9781475763553}, ISSN={0939-6047}, url={http://dx.doi.org/10.1007/978-1-4757-6355-3_10}, DOI={10.1007/978-1-4757-6355-3_10}, booktitle={Random Heterogeneous Materials}, publisher={Springer New York}, author={Torquato, Salvatore}, year={2002}, pages={234–256} }

@article{UliaszMisiak2024,
  title = {Underground Gas Storage in Saline Aquifers: Geological Aspects},
  volume = {17},
  ISSN = {1996-1073},
  url = {http://dx.doi.org/10.3390/en17071666},
  DOI = {10.3390/en17071666},
  number = {7},
  journal = {Energies},
  publisher = {MDPI AG},
  author = {Uliasz-Misiak,  Barbara and Misiak,  Jacek},
  year = {2024},
  month = mar,
  pages = {1666}
}

@article{Vitali2012,
	title = {An extended Eulerian method for contacts in Godunov formulations},
	volume = {92},
	ISSN = {1097-0207},
	url = {http://dx.doi.org/10.1002/nme.4379},
	DOI = {10.1002/nme.4379},
	number = {13},
	journal = {International Journal for Numerical Methods in Engineering},
	publisher = {Wiley},
	author = {Vitali,  E. and Lomov,  I. N. and Antoun,  T. H. and Fujino,  D. H.},
	year = {2012},
	month = jul,
	pages = {1139–1156}
}

@book{Vorobiev2021,
	title = {Geodyn Material Library: Pseudocap models for dry porous rocks},
	url = {http://dx.doi.org/10.2172/1959560},
	DOI = {10.2172/1959560},
	institution = {Office of Scientific and Technical Information (OSTI)},
	author = {Vorobiev,  Oleg},
	year = {2021},
	month = jan, 
	publisher = {Lawrence Livermore National Laboratory. LLNL-TR-818095.}
}

@article{Wang2021,
  title = {Triaxial testing on water permeability evolution of fractured shale},
  volume = {8},
  ISSN = {2054-5703},
  url = {http://dx.doi.org/10.1098/rsos.211270},
  DOI = {10.1098/rsos.211270},
  number = {12},
  journal = {Royal Society Open Science},
  publisher = {The Royal Society},
  author = {Wang,  Menglai and Zhang,  Dongming},
  year = {2021},
  month = dec 
}

@article{Wu2025,
  title = {Advances in Crack Formation Mechanisms,  Evaluation Models,  and Compositional Strategies for Additively Manufactured Nickel-Based Superalloys},
  volume = {143},
  ISSN = {1526-1506},
  url = {http://dx.doi.org/10.32604/cmes.2025.064854},
  DOI = {10.32604/cmes.2025.064854},
  number = {3},
  journal = {Computer Modeling in Engineering \& Sciences},
  publisher = {Tech Science Press},
  author = {Wu,  Huabo and Zhou,  Jialiao and Huang,  Lan and Wang,  Zi and Tan,  Liming and Lv,  Jin and Liu,  Feng},
  year = {2025},
  pages = {2675–2709}
}

@article{Zhang2014,
	title = {Relative Permeability of Coal: A Review},
	volume = {106},
	ISSN = {1573-1634},
	url = {http://dx.doi.org/10.1007/s11242-014-0414-4},
	DOI = {10.1007/s11242-014-0414-4},
	number = {3},
	journal = {Transport in Porous Media},
	publisher = {Springer Science and Business Media LLC},
	author = {Zhang,  Jiyuan and Feng,  Qihong and Zhang,  Xianmin and Wen,  Shengming and Zhai,  Yuyang},
	year = {2014},
	month = dec,
	pages = {563–594}
}

@article{Zhang2016b,
  title = {Crack nucleation using combined crystal plasticity modelling,  high-resolution digital image correlation and high-resolution electron backscatter diffraction in a superalloy containing non-metallic inclusions under fatigue},
  volume = {472},
  ISSN = {1471-2946},
  url = {http://dx.doi.org/10.1098/rspa.2015.0792},
  DOI = {10.1098/rspa.2015.0792},
  number = {2189},
  journal = {Proceedings of the Royal Society A: Mathematical,  Physical and Engineering Sciences},
  publisher = {The Royal Society},
  author = {Zhang,  Tiantian and Jiang,  Jun and Britton,  Ben and Shollock,  Barbara and Dunne,  Fionn},
  year = {2016},
  month = may,
  pages = {20150792}
}

@article{Zhu1999, 
	title={Network modeling of the evolution of permeability and dilatancy in compact rock}, 
	volume={104}, 
	ISSN={0148-0227}, 
	url={http://dx.doi.org/10.1029/1998JB900062}, 
	DOI={10.1029/1998jb900062}, number={B2}, 
	journal={Journal of Geophysical Research: Solid Earth}, 
	publisher={American Geophysical Union (AGU)}, 
	author={Zhu, Wenlu and Wong, Teng‐fong}, 
	year={1999}, 
	month=feb, 
	pages={2963–2971} 
}

@article{Zuo2006, 
	title={A rate-dependent damage model for brittle materials based on the dominant crack}, 
	volume={43}, 
	ISSN={0020-7683}, 
	url={http://dx.doi.org/10.1016/j.ijsolstr.2005.06.083}, 
	DOI={10.1016/j.ijsolstr.2005.06.083}, 
	number={11–12}, 
	journal={International Journal of Solids and Structures}, 
	publisher={Elsevier BV}, 
	author={Zuo, Q.H. and Addessio, F.L. and Dienes, J.K. and Lewis, M.W.}, 
	year={2006},
	month=jun, 
	pages={3350–3380}
}

@inbook{Zhang_2014, title={Microfluidics and Micro Total Analytical Systems}, ISBN={9781455776313}, url={http://dx.doi.org/10.1016/B978-1-4557-7631-3.00003-X}, DOI={10.1016/b978-1-4557-7631-3.00003-x}, booktitle={Molecular Sensors and Nanodevices}, publisher={Elsevier}, author={Zhang, John X.J. and Hoshino, Kazunori}, year={2014}, pages={103–168} }

@article{Zhang2016,
	title = {The stress–strain–permeability behaviour of clay rock during damage and recompaction},
	volume = {8},
	ISSN = {1674-7755},
	url = {http://dx.doi.org/10.1016/j.jrmge.2015.10.001},
	DOI = {10.1016/j.jrmge.2015.10.001},
	number = {1},
	journal = {Journal of Rock Mechanics and Geotechnical Engineering},
	publisher = {Elsevier BV},
	author = {Zhang,  Chun-Liang},
	year = {2016},
	month = feb,
	pages = {16–26}
}

@article{Zhang2022,
	title = {A Model of Stress-Damage-Permeability Relationship of Weakly Cemented Rocks under Triaxial Compressive Conditions},
	volume = {16},
	ISSN = {1996-1944},
	url = {http://dx.doi.org/10.3390/ma16010210},
	DOI = {10.3390/ma16010210},
	number = {1},
	journal = {Materials},
	publisher = {MDPI AG},
	author = {Zhang,  Shizhong and Fan,  Gangwei and Zhang,  Dongsheng and Li,  Wenping and Luo,  Tao and Liang,  Shuaishuai and Fan,  Zhanglei},
	year = {2022},
	month = dec,
	pages = {210}
}

@article{Zhang2024b,
  title = {Permeability heterogeneity effects on density-driven CO2 natural convection and carbon sequestration efficiency},
  volume = {363},
  ISSN = {0016-2361},
  url = {http://dx.doi.org/10.1016/j.fuel.2024.130871},
  DOI = {10.1016/j.fuel.2024.130871},
  journal = {Fuel},
  publisher = {Elsevier BV},
  author = {Zhang,  Qi and Xu,  Quan and Yang,  Yongfei and Iglauer,  Stefan and Liu,  Jie and Liu,  Fugui and Zhang,  Lei and Sun,  Hai and Zhang,  Kai and Yao,  Jun},
  year = {2024},
  month = may,
  pages = {130871}
}

@article{Zhao2020,
  title = {Improved pore network models to simulate single-phase flow in porous media by coupling with lattice Boltzmann method},
  volume = {145},
  ISSN = {0309-1708},
  url = {http://dx.doi.org/10.1016/j.advwatres.2020.103738},
  DOI = {10.1016/j.advwatres.2020.103738},
  journal = {Advances in Water Resources},
  publisher = {Elsevier BV},
  author = {Zhao,  Jianlin and Qin,  Feifei and Derome,  Dominique and Kang,  Qinjun and Carmeliet,  Jan},
  year = {2020},
  month = nov,
  pages = {103738}
}

@article{Zhao2021,
  title = {Study on the technology of enhancing permeability by deep hole presplitting blasting in Sanyuan coal mine},
  volume = {11},
  ISSN = {2045-2322},
  url = {http://dx.doi.org/10.1038/s41598-021-98922-9},
  DOI = {10.1038/s41598-021-98922-9},
  number = {1},
  journal = {Scientific Reports},
  publisher = {Springer Science and Business Media LLC},
  author = {Zhao,  Dan and Wang,  Mingyu and Gao,  Xinhao},
  year = {2021},
  month = oct 
}

@article{Zhu1997,
	title = {Shear-enhanced compaction and permeability reduction: Triaxial extension tests on porous sandstone},
	volume = {25},
	ISSN = {0167-6636},
	url = {http://dx.doi.org/10.1016/S0167-6636(97)00011-2},
	DOI = {10.1016/s0167-6636(97)00011-2},
	number = {3},
	journal = {Mechanics of Materials},
	publisher = {Elsevier BV},
	author = {Zhu,  Wenlu and Montesi,  Laurent G.J. and Wong,  Teng-fong},
	year = {1997},
	month = apr,
	pages = {199–214}
}

\end{document}